# A quantitative method for benchmarking fair income distribution


**Thitithep Sitthiyot[1] and Kanyarat Holasut[2]**

[1] Corresponding Author, Department of Banking and Finance, Faculty of Commerce and Accountancy, Chulalongkorn University, Mahitaladhibesra Bld., 10th Fl., Phayathai Rd., Pathumwan, Bangkok 10330, Thailand, E-mail address: thitithep@cbs.chula.ac.th, ORCID ID: https://orcid.org/0000-0001-9610-2279.

[2] Department of Chemical Engineering, Faculty of Engineering, Khon Kaen University, Mittapap Rd., Muang District, Khon Kaen 40002, Thailand, E-mail address: kanyarat@kku.ac.th, ORCID ID: http://orcid.org/0000-0003-3973-7782.



## Abstract

Concern about income inequality has become prominent in public discourse around the world. However, studies in behavioral economics and psychology have consistently shown that people prefer not equal but fair income distributions. Thus, finding a benchmark that could be used to measure fair income distribution across countries is a theoretical and practical challenge. Here a method for benchmarking fair income distribution is introduced. The benchmark is constructed based on the concepts of procedural justice, distributive justice, and authority's power in professional sports where it is widely agreed as an international norm that the allocations of athlete's salary are outcomes of fair rules, individual and/or team performance, and luck in line with no-envy principle of fair allocation. Using the World Bank data, this study demonstrates how the benchmark could be used to quantitatively gauge whether, for a given value of the Gini index, the income shares by quintile of a country are the fair shares or not, and if not, what fair income shares by quintile of that country should be. Knowing this could be




useful for those involved in setting targets for the Gini index and the fair income shares that are appropriate for the context of each country before formulating policies towards achieving the Sustainable Development Goal 10 and other related SDGs.

# 1. Introduction

Concern about inequality in income distribution has become prominent in public discourse around the world (Mayhew & Wills, 2019). According to the United Nations (2020a), in all countries with data, during the period 2012 to 2017, income is increasingly concentrated at the top where the richest 10% of the population receive at least 20% of the overall income while the bottom 40% receive less than 25% of total income. As income inequality increases, self-reported happiness diminishes, especially among the bottom 40% of the income earners (Oishi et al., 2011). In addition, income inequality predicts a greater degree of violence, obesity, teenage pregnancy, and interpersonal distrust (Wilkinson & Pickett, 2009). In the United States of America, areas with high income inequality tend to have higher divorce and bankruptcy rates than those with more equal income distributions (Frank et al., 2014) and they also suffer from higher homicide rates (Daly et al., 2001). According to the United Nations (2020b), high and rising inequality hinders progress towards the Sustainable Development Goals (SDGs). Reducing inequality within and among countries is regarded as one of the central themes of the 2030 Agenda for Sustainable Development. Target 10.1 of the SDG 10 calls for actions to reduce income-based inequality within countries by progressively achieving and sustaining income growth of the bottom 40% of the population at a rate higher than the national average by 2030 (the United Nations, 2020b).

While there is immense concern about income inequality, both among scholarly community and in general public, and many insist that equality is an important social goal (Starmans et al., 2017), research in behavioral economics, studying distributive behavior in situations where people have earned money being distributed, have found that the majority of people accept income inequalities as fair if the inequalities correspond to differences in



contributions (Cappelen et al., 2014; Almås et al., 2020). Research in political psychology have also found that when people are asked about the ideal distribution of wealth in their country, they actually prefer unequal societies (Norton & Ariely, 2011; Kiatpongsan & Norton, 2014; Norton et al., 2014). Moreover, laboratory studies, cross-cultural research, and experiments with babies and young children have consistently shown that humans naturally favor fair distributions, not equal ones, and that when facing a choice between fairness and equality, people prefer *fair inequality to unfair equality* [emphasis added] (Starmans et al., 2017).

There are two concepts that are commonly accepted which could be used to justify whether or not inequality is seen as fair. The first concept states that people have to agree that inequality results from fair processes with regard to resource allocations (Trump, 2020). This is known as procedural justice where fairness is associated with processes by which the authorities enact rules, resolve disputes, and allocate resources (van Dijke et al., 2019). As long as people generally believe that the outcomes flow from fair allocation procedures, discontent would be muted (Tyler, 2011). Focusing on procedural justice is insufficient to justify whether inequality is viewed as fair, however. It is also relevant to consider the second concept which concerns how benefits and costs are distributed across individuals and/or groups. This is known as distributive justice where fairness is associated with the extent to which the outcomes of processes that allocate benefits and costs are perceived as matching distributive norms such as the equity rule which requires that individuals and/or groups should receive benefits and costs proportional to their contributions (van Dijke et al., 2019). According to van Dijke et al. (2019), studies have shown that people who perceive the distribution of outcomes across individuals, groups, and society as a whole as fair would react in more positive ways. Although inequalities produced through individual effort and merit are seen as fairer than those produced through luck (Almås et al., 2020), both adults and



children agree that inequalities produced by using random allocation devices are acceptable (Karen & Teigen, 2010; Kimbrough et al., 2014; Shaw & Olson, 2014; Choshen-Hillel et al., 2015; Molina et al., 2019). In real life, people often disagree on what is fair because they disagree on whether individual achievements, luck, and efficiency considerations of what maximizes total benefits can justify inequalities (Almås et al., 2010). Thus, what considered as fair do vary across cultures and societies (Rochat et al., 2009; Schäfer et al., 2015; Almås et al., 2020; Apel, 2020). Focusing on the issue of fair income distribution, finding a quantitative benchmark that could be used as a reference for comparing income distribution across countries is therefore a theoretical and practical challenge. According to the Office of the United Nations High Commissioner for Human Rights (OHCHR) (2012), a meaningful benchmark should have a general consensus or an international norm on the choice of a candidate to be used for assessment.

Given the growing concern about income inequality among academic scholars and in general public on the one hand and the empirical evidence showing people's preference for fair income distribution, not equal one on the other, it is important to find out whether the existing income distribution of a country is fair or not. If not, what should the fair distribution of income among population in that country be? While the political economy of what levels of income inequalities that countries would accept as fair remains an uncharted territory (Rogoff, 2012), to our knowledge, there have been very few studies that attempt to quantitatively measure how much inequality in income distribution is fair. These studies are Venkatasubramanian (2009; 2010; 2019), Venkatasubramanian et al. (2015), and Park & Kim (2021). Venkatasubramanian (2009; 2010; 2019) and Venkatasubramanian et al. (2015) develop a benchmark that is based primarily on the concept of maximum entropy in statistical mechanics and information theory, and propose that, in an ideal free market, fair income allocation should follow a log-normal



distribution whereas Park & Kim (2021) propose the concept of a feasible income equality that maximizes the total social welfare and demonstrate that an optimal income distribution that represents the feasible equality could be modeled by using the sigmoid welfare function and the Boltzmann income distribution.

In this study, we introduce an alternative method for benchmarking fair income distribution. Our method is based on the concepts of procedural justice and distributive justice with regard to the allocations of athlete's salary in the realm of professional sports. The justifications for choosing the allocations of athlete's salary in professional sports are mainly because, according to Rogoff (2012), Brannon (2014), Kay (2015, p. 254), and Pritchett & Tiryakian (2020a; 2020b), it is widely agreed as an international norm that the allocations of athlete's salary in professional sports are resulted from fair and transparent rules, individual and/or team effort, as well as luck, all of which fit well with the notions of procedural justice and distributive justice as discussed above. Professional sports are generally considered as a universal language since the rules of the games, the performances of the athletes, and the outcomes of the competitions are, by and large, understandable to and could be viewed and judged by ordinary people as fair, regardless of their origin, background, religious belief, or economic status. Brannon (2014) and Pritchett & Tiryakian (2020a; 2020b) note that few fans seem to begrudge the professional athletes for their money. This stylized fact has long been known in the fairness literature as no-envy principle of fair allocation which was introduced by the Nobel prize laureate in economics Jan Tinbergen in his article published in 1930. In line with no-envy principle of fair allocation, Harvard economist Kenneth Rogoff expresses his amazement over the public's blasé acceptance of the salaries of sports stars compared to its low regard for superstars in business and finance (Rogoff, 2012). This is because it is easier to grasp



the sacrifices an athlete in professional sport made to achieve stardom than a successful chief executive officer where vilification runs rampant (Brannon, 2014). The United Nations also recognizes the contribution of sports to the SDGs (the United Nations Office on Sport Development and Peace (UNOSDP), 2018). As stated by the UNOSDP (2018), sport value such as fairness can serve as an example for an economic system that builds on fair competition. In addition to the ideas of procedural justice and distributive justice, we follow van Dijke et al. (2019) by including the idea of power of authority as a basis to support our selection of professional sports for developing our method. van Dijke et al. (2019) define power of authority as a capacity to detect and punish misconducts and rule violations. According to van Dijke et al. (2019), high authority's power is an important element of the process underlying the effects of procedural justice and distributive justice on voluntary rule compliance. This is similar to professional sports. When the power of sport authority is high, the athlete tends to voluntarily comply and adheres to the rules of the game.

Viewing through the lens of procedural justice, distributive justice, and authority's power in professional sports, the distribution of athlete's salary could thus be justified as a plausible candidate for developing a neutral benchmark for measuring fair income distribution since the salary is allocated based on the performance of each individual athlete (distributive justice) who competes according to fair and transparent rules (procedural justice) that are written and administered by the sports authority (power of authority), all of which, by and large, are comprehensible to and could be observed and judged by ordinary people around the world as fair as pointed out by Rogoff (2012), Brannon (2014), Kay (2015), the UNOSDP (2018), and Pritchett & Tiryakian (2020a; 2020b), and hence satisfies both the no-envy principle of fair allocation (Tinbergen, 1930) and the general consensus or the international norm criterion of a meaningful



benchmark according to the OHCHR (2012). While this study proposes the use of the distribution of professional athlete's salary for developing a benchmark for measuring fair income distribution, we are fully aware that other types of data could also be used if they are considered as fairer and more universally accepted than the data on the distribution of athlete's salary in professional sports, provided such data can be numerically ranked.

To develop our benchmark, the data on the annual salaries of athletes from 11 well-known professional sports in 2019 or the latest from Sitthiyot (2021) comprising 6,709 observations are employed in order to construct the fairness lines. These fairness lines are derived from the statistical relationship between the salary shares of the athlete in each quintile and the Gini index for the athlete's salary of 11 professional sports. They are the Women's National Basketball Association (WNBA), the English Premier League (EPL), the National Football League (NFL), the National Hockey League (NHL), the Major League Baseball (MLB), the National Basketball Association (NBA), the Professional Golfers' Association of America (PGA), the Ladies Professional Golf Association (LPGA), the Major League Soccer (MLS), the Association of Tennis Professionals (ATP), and the Women's Tennis Association (WTA).

It is very important to note that rules, regulations, and player's codes of conduct in these professional sports are written and administered by relevant international sports authorities whose members are from different countries. In addition, the athletes competing in these professional sports come from diverse ethnic, social, national, and regional backgrounds and receive compensations based on their performances, skills, and luck. Furthermore, these professional sports tournaments, by and large, are taken place in different countries around the world. For these reasons, employing the data on the athlete's salary from 11 professional sports



should avoid or mitigate the problem of selection bias since they do not come from one specific group of population and/or one particular country.

Using the annual data on the Gini index and the income shares by quintile in 2015 from the World Bank (2020a) containing 75 countries, this study shows how the fairness lines could be used as a benchmark to quantitatively gauge whether, for an existing value of a country's Gini index, the income shares in each quintile of that country are the fair shares or not, and if not, what level of fair income shares in each quintile of that country should be. We also demonstrate that, by using our fairness benchmark, policymakers could choose a combination of the Gini index and the fair income shares by quintile that are appropriate for the context of that country. This could benefit whoever involved in formulating strategies and policies aimed to achieve not equal, but fair income distribution societies, the SDG 10, and other related SDGs. In addition, given that the fair income distribution depends on the country context and can vary considerably, our method could be used to compare fair income distribution among different groups within a country as well as across countries since they are measured using the same benchmark.

## 2. Materials and Methods

For each of 11 professional sports, namely, WNBA, EPL, NFL, NHL, MLB, NBA, PGA, LPGA, MLS, ATP, and WTA, the derivation of the benchmark for fair income distribution begins by normalizing the annual salaries of the athlete and ranking them in an ascending order. We then plot the Cartesian coordinates where the abscissa is the cumulative normalized rank of athlete's salary (x) and the ordinate is the cumulative normalized athlete's salary (y). This would give us the actual Lorenz plot for each type of sports. Next, we fit the estimated Lorenz curve to the actual Lorenz plot for each type of 11 professional sports by employing a parametric



functional form which has a closed-form expression for the Gini index developed by Sitthiyot &

Holasut (2021). This specified functional form is constructed based on the weighted average of 2

well-known functional forms for estimating the Lorenz curve which are the exponential function

$(y(x) = x^P)$ and the functional form implied by Pareto distribution $(y(x) = 1 - (1-x)^{\frac{1}{P}})$. Note

that, if separately integrating the exponential function and the functional form implied by Pareto

distribution from 0 to 1, it can be seen that both functional forms have the same area under the

Lorenz curve which is equal to $\frac{1}{P+1}$. By assigning the weight $(1-k)$ to the exponential function

and the weight k to the functional form implied by Pareto distribution, the specified functional

form based on the weighted average of these 2 well-known functional forms is as follows:

$$y(x) = [1-k] * x^P + k * [1 - (1-x)^{\frac{1}{P}}] \tag{1}$$

$$0 \leq k \leq 1$$

$$1 \leq P$$

This specified functional form satisfies all required properties of the Lorenz curve in that

it must be a monotonically increasing function $\left(\frac{dy}{dx} \geq 0, \frac{d^2y}{dx^2} \geq 0\right)$ and must pass two coordinates

which are (0, 0) and (1, 1) as illustrated in Figure 1.



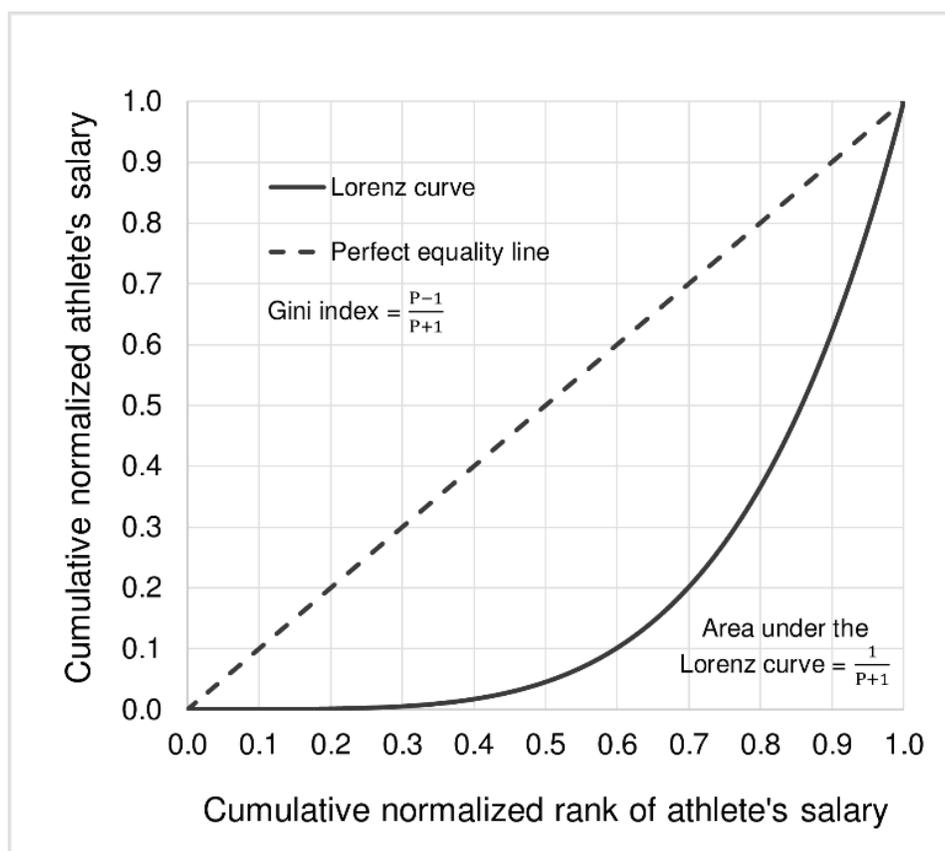

**Figure 1. The Lorenz curve and the Gini index.**

According to Sitthiyot & Holasut (2021), the parameter k is the weight that controls the curvature of the estimated Lorenz curve such that the Gini index remains constant, whereas the parameter P represents the degree of inequality in the distribution of the athlete's salary. The values of parameters k and P can vary depending upon the curvature of the estimated Lorenz curve and the degree of inequality in the distribution of the athlete's salary for each professional sport. The estimated Lorenz curve based on equation (1) would be used to calculate the Gini index for the athlete's salary which measures inequality in the distribution of salary of the athlete for each type of 11 professional sports. The Gini index takes values between 0 and 1. The closer the index is to 0, the more equal the distribution of the athlete's salary. The closer the index is to 1, the more



unequal the distribution of the athlete's salary. Sitthiyot & Holasut (2020) note that the advantage of the Gini index as a measure of income inequality is that the inequality of the entire income distribution could be summarized by using a single number that is easy to interpret since its values are bounded between 0 and 1. This would allow for comparison among countries with different population sizes. Moreover, the data on the Gini index is easy to access, regularly updated and published by countries and/or international organizations. For these reasons, the Gini index has arguably been the most popular measure of inequality despite the fact that there are well over 50 inequality indices as discussed in Coulter (1989).

Given the estimated Lorenz curve for each of 11 professional sports, we can calculate the area under the curve by integrating the estimated Lorenz equation from 0 to 1 $\left( \int_0^1 y(x)dx \right)$. Based on the parametric functional form proposed by Sitthiyot & Holasut (2021) as shown in equation (1), the area under the curve equals $\frac{1}{P+1}$. After obtaining the area under the estimated Lorenz curve, the Gini index for the athlete's salary for each type of 11 professional sports can be calculated as $1 - 2 * \int_0^1 y(x)dx$ which is equal to $\frac{P-1}{P+1}$. Note that when $P = 1$, the Gini index $= 0$, which implies perfect equality (all athletes in the same professional sports receive the same salary share). When $P = \infty$, the Gini index $= 1$, which implies perfect inequality (one athlete has all the salary and the rest has none for the same professional sports).

In addition to the Gini index for the athlete's salary, the estimated Lorenz equation can be used to compute the salary shares by quintile for each of 11 professional sports. Thus, each of 11 professional sports would have its own Gini index and the salary shares by quintile where the 1st quintile is the salary share held by the lowest 20% (0% ≤ salary share ≤ 20%), the 2nd quintile is the salary share held by the second 20% (20% < salary share ≤ 40%), the 3rd quintile is the salary share held by the third 20% (40% < salary share ≤ 60%), the 4th quintile is the salary share held by



the fourth 20% (60% < salary share ≤ 80%), and the 5th quintile is the salary share held by the top

20% (80% < salary share ≤ 100%).

Provided that the distributions of the athlete's salary of all 11 professional sports satisfy the

notions of fairness according to the concepts of procedural justice and distributive justice together

with the concept of power of authority which reflect the no-envy principle of fair allocation

introduced by Tinbergen (1930) and the general consensus or the international norm criterion for a

meaningful benchmark according to the OHCHR (2012) as discussed in the Introduction section,

the next step is to construct the fairness lines. This could be done by estimating the statistical

relationship between the salary shares of the athlete in each quintile ($Q^S_i$, i = 1, 2, …, 5) and the

Gini index for the athlete's salary ($Gini_S$) using the data from each type of 11 professional sports.

Note that, for both theoretical and practical purposes with regard to the mathematical properties of

the Lorenz curve and those of the Gini index, we impose five conditions when estimating the

relationship between the salary shares of the athlete in each quintile ($Q^S_i$, i = 1, 2, …, 5) and the

Gini index for the athlete's salary ($Gini_S$). The first condition is that the fairness line for the 5th

quintile must pass 2 coordinates which are (0, 0.2) and (1, 1). The justification for the first

condition is that when the Gini index for the athlete's salary is equal to 0, the salary share in the 5th

quintile must be equal to 0.2, and when the Gini index for the athlete's salary is equal to 1, the

salary share in the 5th quintile must be equal to 1. For the second condition, the fairness line for

other 4 quintiles must pass 2 coordinates which are (0, 0.2) and (1, 0). The rationale for the second

condition is that when the Gini index for the athlete's salary equals 0, the salary shares in the 1st,

the 2nd, the 3rd, and the 4th quintile all have to be equal to 0.2. When the Gini index for the athlete's

salary is equal to 1, the salary shares in the 1st, the 2nd, the 3rd, and the 4th quintile all have to be

equal to 0. Given the property of the Lorenz curve must be a monotonically increasing function,



the third condition is that, for a given value of the Gini index for the athlete's salary, $0 \leq$ the salary share in the 1st quintile ($Q^S_1$) $\leq$ the salary share in the 2nd quintile ($Q^S_2$) $\leq$ the salary share in the 3rd quintile ($Q^S_3$) $\leq$ the salary share in the 4th quintile ($Q^S_4$) $\leq$ the salary share in the 5th quintile ($Q^S_5$) $\leq$ 1. For the fourth condition, $\frac{dQ^S_1}{dGini_s}\Big|_{Gini_S=0} \leq \frac{dQ^S_2}{dGini_s}\Big|_{Gini_S=0} \leq \frac{dQ^S_3}{dGini_s}\Big|_{Gini_S=0} \leq \frac{dQ^S_4}{dGini_s}\Big|_{Gini_S=0} \leq$ $\frac{dQ^S_5}{dGini_s}\Big|_{Gini_S=0}$. That is when the Gini index for the athlete's salary = 0, the slope of the estimated fairness line for the 1st quintile $\leq$ the slope of the estimated fairness line for the 2nd quintile $\leq$ the slope of the estimated fairness line for the 3rd quintile $\leq$ the slope of the estimated fairness line for the 4th quintile $\leq$ the slope of the estimated fairness line for the 5th quintile. Regarding the fifth condition, the summation of the salary shares by quintile ($\sum_{i=1}^{5} Q^S_i$) for each professional sport must equal 1, for a given value of the Gini index for the athlete's salary. Given these five conditions, the curve fitting technique based on minimizing error sum of squares is applied in order to find the statistical relationship between the salary shares of the athlete in each quintile ($Q^S_i$, i = 1, 2, …, 5) and the Gini index for the athlete's salary ($Gini_S$). This would give us 5 fairness lines, one for each quintile, that could be used to estimate the fair salary shares in each quintile for a given value of the Gini index for the athlete's salary.

To measure whether the income distribution of a country is fair or not, we plot the Cartesian coordinate where the abscissa is the income Gini index and the ordinate is the income share in each quintile of that country ($Q^I_i$, i = 1, 2, …, 5). We then compare to see whether the ordered-pair (income Gini index, $Q^I_i$) is on, above, or below the fairness line for the same value of the income Gini index. The closer the ordered-pair (income Gini index, $Q^I_i$) of a country is to the fairness line, the fairer the income share in that quintile relative to the professional athlete's salary benchmark. If the ordered-pair (income Gini index, $Q^I_i$) of a country lies above the fairness line, it



indicates that the income earners in that quintile receive income share more than the fair share. In contrast, if the ordered-pair (income Gini index, $Q^I_i$) of a country lies below the fairness line, it suggests that the income earners in that quintile receive income share lower than the fair share. In practice, for an existing value of the income Gini index of a country, the degree of fairness in income shares by quintile could be quantitatively measured by calculating the percentage deviation (in absolute value) of a country's actual income shares in each quintile from the corresponding fair income share based on the professional athlete's salary benchmark. In this way, not only would our method be able to identify which quintile the income share of a country is fair or not fair relative to the benchmark derived from the athlete's salary from 11 professional sports, but also suggest what the fair share of income in each quintile of that country should be for an existing value of the Gini index.

In this study, the data on the annual salaries of the athletes from 11 professional sports in 2019 or the latest from Sitthiyot (2021) comprising 6,709 observations are used in order to construct the fairness lines. The data on the income Gini index and the income shares by quintile in 2015 from the World Bank (2020a) comprising 75 countries are employed in order to demonstrate how the fairness lines could be used as a benchmark for quantitatively measuring whether the income distribution of a country is fair or not. Note that the reason we use the data on the income Gini index and the income shares by quintile in 2015 for demonstrating our method is mainly because there are more observations than those in 2016 and 2017, and no data on the Gini index and income shares by quintile were available for 2018 and 2019 at the time we accessed the World Bank website. Table 1 reports the descriptive statistics of the athlete's salary from 11 professional sports while the descriptive statistics of the income Gini index and the income shares by quintile are provided in Table 2.

**Table 1. The descriptive statistics of the athletes' salaries from 11 professional sports.** The unit of currency is the United States dollar except for the EPL where the unit of currency is the Pound Sterling.

| Type of sports | Mean | Median | Mode | Minimum | Maximum | Standard deviation | No. of players |
|---|---|---|---|---|---|---|---|
| 1. WNBA | 73,190.60 | 59,718 | 115,000 | 2,723 | 127,500 | 32,589.64 | 151 |
| 2. EPL | 2,695,133.83 | 1,976,000 | 1,820,000 | 0.00 | 19,500,000 | 2,489,890.55 | 523 |
| 3. NFL | 4,682,534.29 | 3,000,000 | 3,095,000 | 831,349 | 30,700,000 | 4,525,065.24 | 1,000 |
| 4. NHL | 2,618,049.97 | 1,237,500 | 700,000 | 675,000 | 12,500,000 | 2,396,904.17 | 1,000 |
| 5. MLB | 7,985,791.19 | 5,000,000 | 4,000,000 | 583,500 | 37,666,666 | 7,708,701.88 | 481 |
| 6. NBA | 7,600,037.45 | 3,500,000 | 1,620,564 | 208,509 | 40,231,758 | 8,767,208.31 | 517 |
| 7. PGA | 1,235,495.42 | 838,030 | - | 5,910 | 9,684,006 | 1,433,077.03 | 264 |
| 8. LPGA | 39,064.90 | 21,380 | 10,874 | 4,015 | 313,272 | 58,022.09 | 77 |
| 9. MLS | 409,288.28 | 175,135 | 70,250 | 56,250 | 7,200,000 | 718,621.46 | 658 |
| 10. ATP | 37,474.00 | 1,084 | 54 | 54 | 3,915,011 | 158,294.82 | 1,070 |
| 11. WTA | 27,520.67 | 625 | 147 | 37 | 2,916,508 | 122,838.31 | 968 |

**Table 2. The descriptive statistics of the income Gini index and the income shares by quintile (in decimals) of 75 countries.**

| Indicator | Mean | Median | Mode | Minimum | Maximum | Standard deviation | No. of countries |
|---|---|---|---|---|---|---|---|
| Income Gini index | 0.369 | 0.356 | 0.318 | 0.254 | 0.591 | 0.081 | 75 |
| Income share held by the top 20% | 0.441 | 0.421 | 0.384 | 0.350 | 0.637 | 0.066 | 75 |
| Income share held by the fourth 20% | 0.219 | 0.221 | 0.226 | 0.179 | 0.247 | 0.012 | 75 |
| Income share held by the third 20% | 0.158 | 0.163 | 0.175 | 0.098 | 0.187 | 0.019 | 75 |
| Income share held by the second 20% | 0.115 | 0.118 | 0.121 | 0.058 | 0.147 | 0.021 | 75 |
| Income share held by the lowest 20% | 0.068 | 0.071 | 0.078 | 0.028 | 0.100 | 0.019 | 75 |

**3. Results**

We first report the values of the parameters k and P as well as the values of coefficient of determination ($R^2$) for the estimated Lorenz equations for each type of 11 professional sports which are WNBA, EPL, NFL, NHL, MLB, NBA, PGA, LPGA, MLS, ATP, and WTA. The values of k and P are estimated by using equation (1) and the curve fitting technique based on minimizing error sum of squares as described in the Materials and Methods section. They are shown in Table 3.

**Table 3. The values of parameter k and P as well as $R^2$ from the estimated Lorenz curves for 11 types of professional sports.**

| Type of sports | k | P | $R^2$ |
|---|---|---|---|
| 1. WNBA | 0.21 | 1.66 | 0.9976 |
| 2. EPL | 0.39 | 2.62 | 0.9995 |
| 3. NFL | 0.50 | 2.76 | 0.9999 |
| 4. NHL | 0.48 | 2.76 | 0.9990 |
| 5. MLB | 0.28 | 2.96 | 0.9986 |
| 6. NBA | 0.31 | 3.52 | 0.9949 |
| 7. PGA | 0.25 | 3.55 | 0.9988 |
| 8. LPGA | 0.48 | 4.04 | 0.9951 |
| 9. MLS | 0.55 | 4.05 | 0.9999 |
| 10. ATP | 0.09 | 14.39 | 0.9999 |
| 11. WTA | 0.08 | 14.69 | 0.9999 |

Note that the estimated Lorenz equations fit the actual Lorenz plots that represent the relationship between the cumulative normalized athlete's salary on the vertical axis and the cumulative normalized rank of the athlete's salary on the horizontal axis for each of 11 professional sports quite well with the values of $R^2$ ranging between 0.9949 and 0.9999.



We then use these estimated Lorenz curves to calculate the values of the Gini index for the athlete's salary (Gini$_S$) and the salary shares by quintile (Q$^S_i$, i = 1, 2, …, 5) (in decimals) for each of 11 professional sports. The results are reported in Table 4.

**Table 4. The values of the Gini index for the athletes' salaries and the salary shares by quintile (in decimals) for each type of the 11 professional sports.** The 2 conditions are also included which are perfect equality and perfect inequality where the Gini index for the athletes' salaries takes the values of 0 and 1, respectively.

| Type of sports | Gini$_S$ | Q$^S_5$ | Q$^S_4$ | Q$^S_3$ | Q$^S_2$ | Q$^S_1$ |
|---|---|---|---|---|---|---|
| Perfect equality | 0.000 | 0.200 | 0.200 | 0.200 | 0.200 | 0.200 |
| 1. WNBA | 0.247 | 0.323 | 0.248 | 0.199 | 0.148 | 0.081 |
| 2. EPL | 0.447 | 0.480 | 0.244 | 0.151 | 0.084 | 0.041 |
| 3. NFL | 0.468 | 0.509 | 0.228 | 0.139 | 0.080 | 0.045 |
| 4. NHL | 0.469 | 0.507 | 0.231 | 0.140 | 0.079 | 0.043 |
| 5. MLB | 0.495 | 0.511 | 0.256 | 0.141 | 0.066 | 0.026 |
| 6. NBA | 0.557 | 0.571 | 0.243 | 0.116 | 0.048 | 0.021 |
| 7. PGA | 0.560 | 0.569 | 0.252 | 0.117 | 0.045 | 0.018 |
| 8. LPGA | 0.603 | 0.631 | 0.205 | 0.094 | 0.043 | 0.027 |
| 9. MLS | 0.604 | 0.637 | 0.195 | 0.092 | 0.046 | 0.030 |
| 10. ATP | 0.870 | 0.954 | 0.040 | 0.003 | 0.002 | 0.001 |
| 11. WTA | 0.873 | 0.957 | 0.038 | 0.003 | 0.002 | 0.001 |
| Perfect inequality | 1.000 | 1.000 | 0.000 | 0.000 | 0.000 | 0.000 |

Excluding the cases of perfect equality and perfect inequality, the results in Table 4 indicate that the degree of inequality in the distributions of athlete's salary, measured by the Gini index, could be as low as 0.247 (WNBA) and as high as 0.873 (WTA). In addition, the salary share of the top 20% (Q$^S_5$) could be as low as 0.323 (WNBA) and as high as 0.957 (WTA) whereas the salary share of the bottom 20% (Q$^S_1$) could be as low as 0.001 (ATP and WTA) and as high as 0.081(WNBA), resulting in the salary gap between the top 20% (Q$^S_5$) and the bottom 20% (Q$^S_1$) to be as low as 3.99 times (WNBA) and as high as 957 times (WTA). These results clearly illustrate that a high degree of inequality in the distribution of salary and a substantial



salary gap between the top 20% and the bottom 20% could be acceptable if they are procedurally and distributively viewed as fair. Table 5 reports the descriptive statistics of the Gini index for the athlete's salary and the salary shares by quintile of 11 professional sports.

**Table 5. The descriptive statistics of the Gini index for the athletes' salaries and the salary shares by quintile (in decimals) of 11 professional sports.** The salary shares when the Gini index = 0.000 and 1.000 are excluded when computing the descriptive statistics.

| Indicator | Mean | Median | Mode | Minimum | Maximum | Standard deviation | No. of sports |
|---|---|---|---|---|---|---|---|
| Gini index for athletes' salaries | 0.5630 | 0.5575 | - | 0.2470 | 0.8725 | 0.1815 | 11 |
| Salary share held by the top 20% | 0.6045 | 0.5690 | - | 0.3233 | 0.9568 | 0.1930 | 11 |
| Salary share held by the fourth 20% | 0.1982 | 0.2310 | - | 0.0377 | 0.2562 | 0.0811 | 11 |
| Salary share held by the third 20% | 0.1086 | 0.1168 | - | 0.0027 | 0.1995 | 0.0600 | 11 |
| Salary share held by the second 20% | 0.0582 | 0.0478 | - | 0.0016 | 0.1476 | 0.0408 | 11 |
| Salary share held by the lowest 20% | 0.0304 | 0.0266 | - | 0.0013 | 0.0813 | 0.0224 | 11 |

The descriptive statistics of the Gini index for the athlete's salary and the salary shares by quintile as reported in Table 5 could be compared to the descriptive statistics of the income Gini index and the income shares by quintile in 2015 as shown in Table 2. Figure 2 shows the plots of the minimum and the maximum values of the Gini index for the athlete's salary and those of the salary shares by quintile vs. the plots of the minimum and the maximum values of the income Gini index and those of the income shares by quintile in 2015.

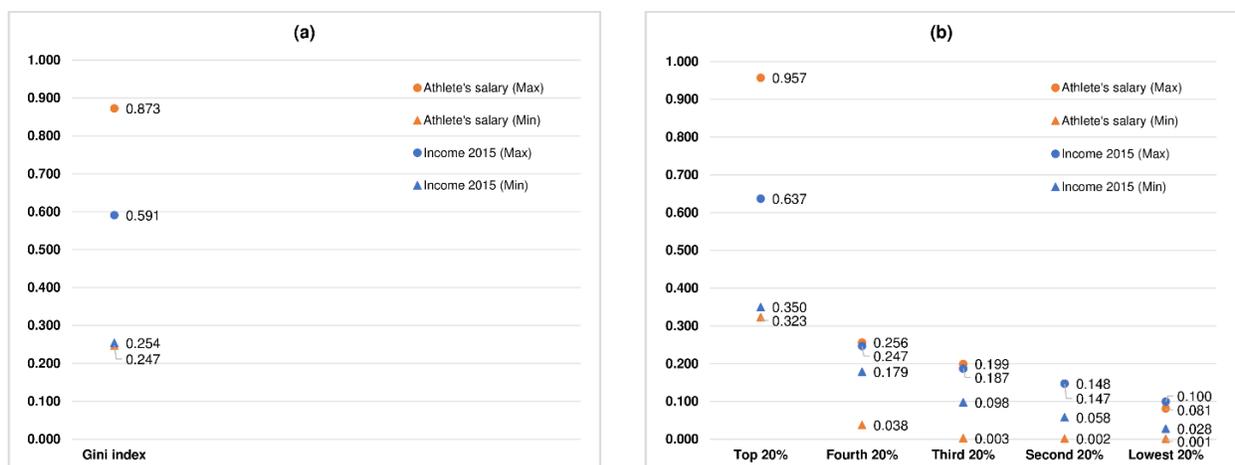

**Figure 2. The minimum and the maximum values of the Gini index for the athlete's salary and those of the salary shares by quintile vs. the minimum and the maximum values of the income Gini index and those of the income shares by quintile in 2015.** (a) Gini index. (b) Income shares by quintile.

It can be seen from Figures 2a and 2b that the minimum and the maximum values of the income Gini index and the income shares by quintile in 2015 lie within the corresponding minimum and maximum values of the Gini index for the athlete's salary and the salary shares by quintile except for the lowest 20% where the maximum value of the income share (0.100) is slightly higher than that of the salary share (0.081). In addition, we collect the data on the income Gini index and the income shares by quintile for the entire period between 1995 and 2015 from the World Bank (2020a) and find their minimum and maximum values. The findings indicate



that the minimum and the maximum values of the income Gini index and the income shares by quintile from 1995 to 2015, by and large, lie within the corresponding minimum and maximum values of the Gini index for the athlete's salary and the salary shares by quintile. Out of 252 observations on the minimum and the maximum values of the income Gini index and the income shares by quintile from 1995 to 2015, there are 9 observations on the income Gini index whose values (0.230-0.246) are slightly below the minimum value of the Gini index for the athlete's salary (0.247) while there are 21 observations on the income share of the lowest 20% whose values (0.090-0.109) are just above the maximum value of the salary share of the lowest 20% (0.081). This is to show that, for the values of the Gini index for the athlete's salary and those of the salary shares by quintile to be used for developing the fairness benchmark, they should, by and large, cover not only the values of the income Gini index and those of the income shares by quintile of 75 countries in the sample but also the values of the income Gini index and those of the income shares by quintile across a wide range of countries and time periods.

Given the allocations of the athlete's salary of all 11 professional sports are resulted from fair and transparent rules written and administered by international sports authorities, individual and/or team performance, as well as luck, all of which correspond to the notions of procedural justice, distributive justice, and authority's power as discussed in the Introduction section, the values of the Gini index for the athlete's salary ($Gini_S$) and the salary shares by quintile ($Q^S_i$, i = 1, 2, …, 5) can thus be used to construct the fairness lines, one for each quintile. This could be done by estimating the statistical relationship between the salary shares of the athlete in each quintile and the Gini index for the athlete's salary based on five conditions which are: 1) the fairness line for the 5th quintile has to pass 2 coordinates, namely, (0, 0.2) and (1, 1); 2) the fairness lines for other 4 quintiles must pass 2 coordinates, namely, (0, 0.2) and (1, 0); 3) for a given value of the



Gini index for the athlete's salary, $0 \leq$ the salary share in the 1st quintile ($Q^S_1$) $\leq$ the salary share in the 2nd quintile ($Q^S_2$) $\leq$ the salary share in the 3rd quintile ($Q^S_3$) $\leq$ the salary share in the 4th quintile ($Q^S_4$) $\leq$ the salary share in the 5th quintile ($Q^S_5$) $\leq 1$; 4) $\frac{dQ^S_1}{dGini_S}\Big|_{Gini_S=0} \leq \frac{dQ^S_2}{dGini_S}\Big|_{Gini_S=0} \leq \frac{dQ^S_3}{dGini_S}\Big|_{Gini_S=0} \leq \frac{dQ^S_4}{dGini_S}\Big|_{Gini_S=0} \leq \frac{dQ^S_5}{dGini_S}\Big|_{Gini_S=0}$; and 5) for a given value of the Gini index for the athlete's salary, the summation of the salary shares by quintile ($\sum_{i=1}^{5} Q^S_i$) for each sports must be equal to 1. These five conditions are imposed in order to guarantee that the estimated fairness lines satisfy the mathematical properties of the Lorenz curve and those of the Gini index. By applying the curve fitting technique based on minimizing sum of squared errors, we find that the estimated equations for the fairness lines can be fitted by the 4th degree polynomial function. They are as follows:

$$Q^S_1 = 0.1956Gini^4_S - 0.6612Gini^3_S + 0.9356Gini^2_S - 0.6700Gini_S + 0.20, R^2 = 0.9836 \qquad (2)$$

$$Q^S_2 = -0.4914Gini^4_S + 1.3968Gini^3_S - 1.1195Gini^2_S + 0.0141Gini_S + 0.20, R^2 = 0.9935 \qquad (3)$$

$$Q^S_3 = 0.2545Gini^4_S + 0.0508Gini^3_S - 0.6650Gini^2_S + 0.1597Gini_S + 0.20, R^2 = 0.9915 \qquad (4)$$

$$Q^S_4 = 1.9481Gini^4_S - 3.3533Gini^3_S + 1.0625Gini^2_S + 0.1428Gini_S + 0.20, R^2 = 0.9737 \qquad (5)$$

$$Q^S_5 = -1.9067Gini^4_S + 2.5669Gini^3_S - 0.2136Gini^2_S + 0.3534Gini_S + 0.20, R^2 = 0.9960 \qquad (6)$$

Figure 3 illustrates the fairness lines representing the relationship between the salary shares of the professional athlete in each quintile and the Gini index for the athlete's salary.



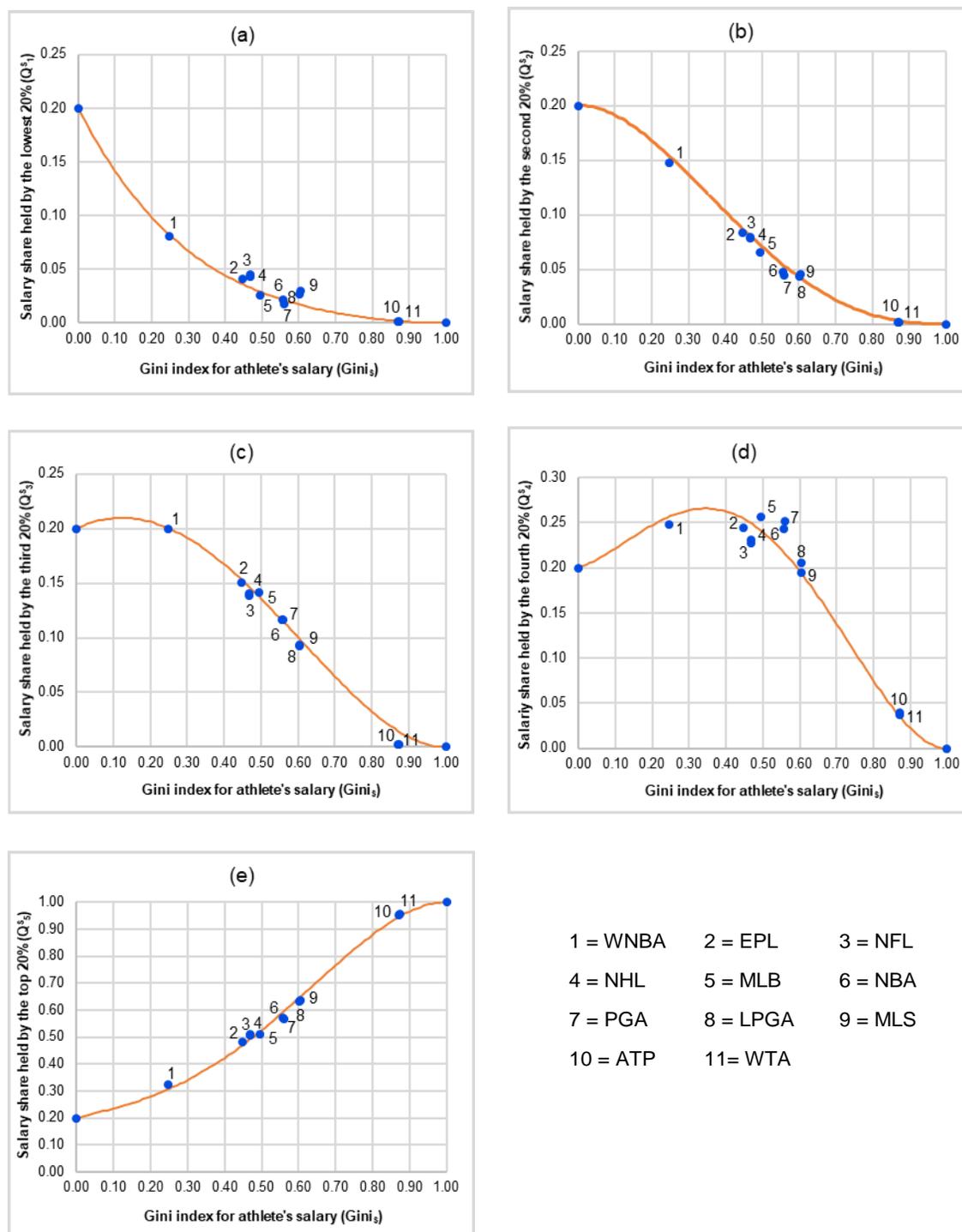

**Figure 3. The estimated fairness lines representing the relationship between the salary shares in each quintile ($Q^S_i$, i = 1, 2, …, 5) and the Gini index for the athlete's salary ($Gini_s$) from 11 professional sports.** (a) Salary share held by the lowest 20% ($Q^S_1$). (b) Salary share held by the second 20% ($Q^S_2$). (c) Salary share held by the third 20% ($Q^S_3$). (d) Salary share held by the fourth 20% ($Q^S_4$). (e) Salary share held by the top 20% ($Q^S_5$).



Next, we show how the fairness lines could be used as a benchmark for measuring whether income shares in each quintile of a country are regarded as fair or not. This could be done by comparing the Cartesian coordinates of each country in the sample where the abscissa is the income Gini index and the ordinate is the income shares by quintile ($Q^I_i$, i = 1, 2, …, 5) to the fairness lines in the corresponding quintiles as shown in Figure 3. The closer the ordered-pair (income Gini index, $Q^I_i$) of a country is to the fairness line, the fairer the income share in that quintile. The further away the ordered-pair (income Gini index, $Q^I_i$) is from the fairness line, the more unfair the income share of a country in that quintile. In this study, we measure the degree of fairness by calculating the percentage deviation (in absolute value) of a country's income share in each quintile from the fair income share based on our benchmark derived from the relationship between the salary shares of the athlete in each quintile and the Gini index for the athlete's salary from 11 professional sports.

Using the data on the income Gini index and the income shares by quintile in 2015 (in decimals) from the World Bank (2020a) covering 75 countries, Figure 4 illustrates the scatter plots of the Cartesian coordinate where the abscissa is the income Gini index and the ordinate is the income shares by quintile of 75 countries in the sample ($Q^I_i$, i = 1, 2, …, 5).



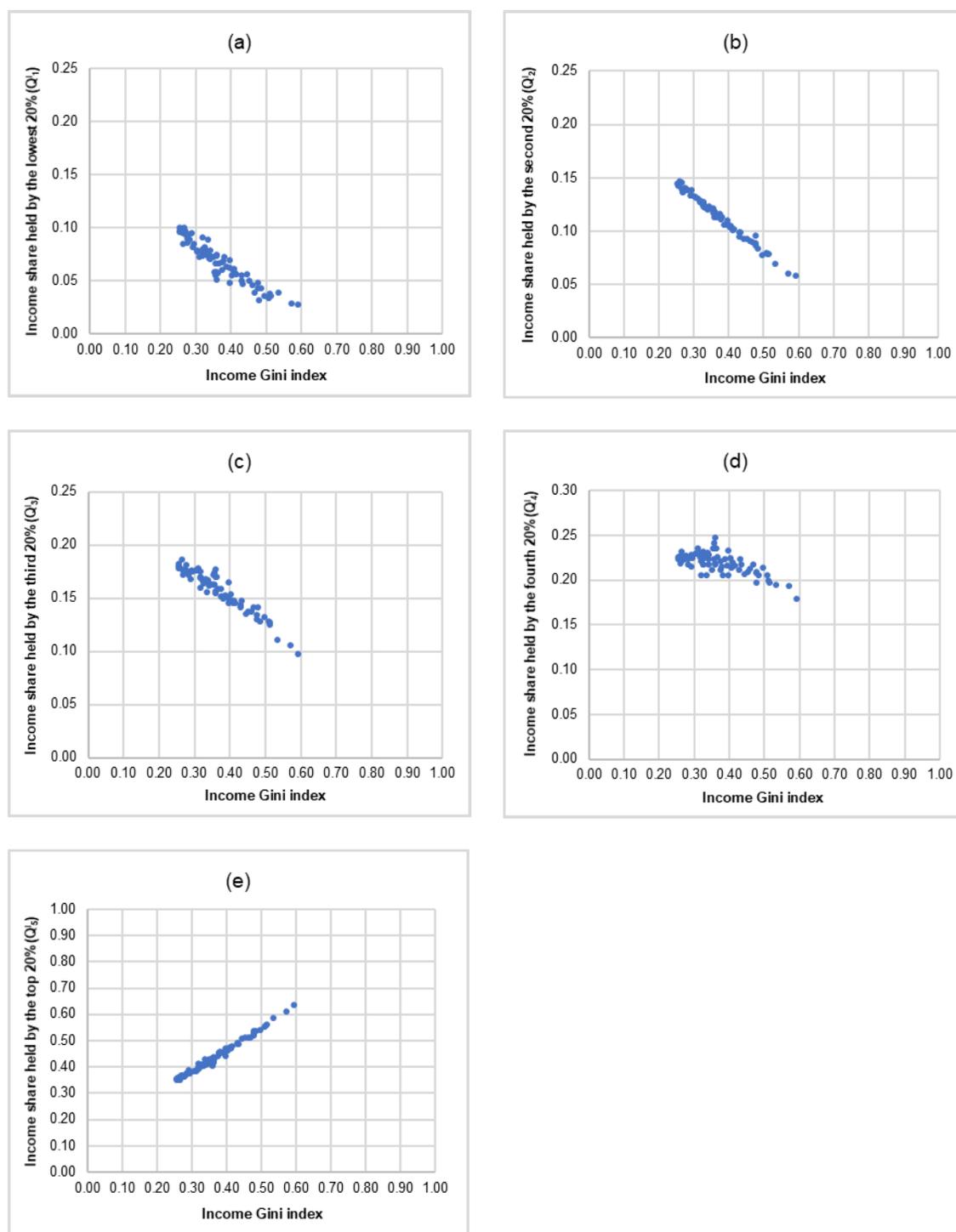

**Figure 4. Scatter plots illustrating the relationship between the income Gini index and the income shares by quintile ($Q^I_i$, i = 1, 2, …, 5) of 75 countries.** (a) Income share held by the lowest 20% ($Q^I_1$). (b) Income share held by the second 20% ($Q^I_2$). (c) Income share held by the third 20% ($Q^I_3$). (d) Income share held by the fourth 20% ($Q^I_4$). (e) Income share held by the top 20% ($Q^I_5$).



It is interesting to note that the scatter plots of the ordered-pairs (income Gini index, $Q^I_i$) of all 75 countries in the corresponding quintile as shown in Figure 4 share similar pattern with the scatter plots of the ordered-pairs $(Gini_S, Q^S_i)$ of 11 professional sports in each quintile used to construct the fairness lines as shown in Figure 3. The scatter plots of the ordered-pairs $(Gini_S, Q^S_i)$ of 11 professional sports in Figure 3 and the scatter plots of the ordered-pairs (income Gini index, $Q^I_i$) of 75 countries in Figure 4 could be viewed together which are shown in Figure 5.



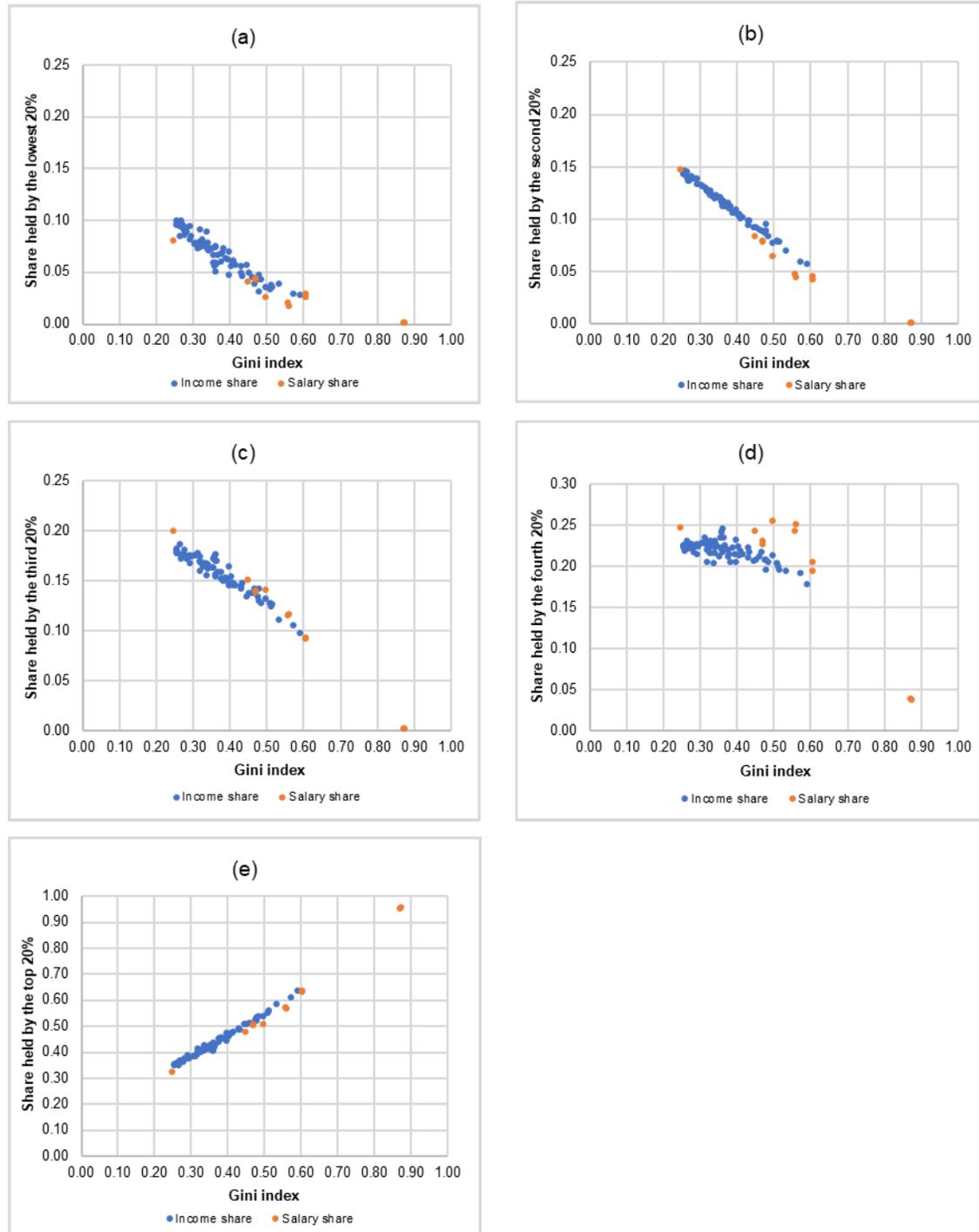

**Figure 5. Similarity between the scatter plots of the ordered-pairs of the income Gini index and the income shares by quintile (Income Gini index, $Q^I_i$) of 75 countries and the scatter plots of the ordered-pairs of the Gini index for the athlete's salary and the salary shares by quintile (Ginis, $Q^S_i$) of 11 professional sports.** (a) Share held by the lowest 20%. (b) Share held by the second 20%. (c) Share held by the third 20%. (d) Share held by the fourth 20%. (e) Share held by the top 20%.



Figure 6 illustrates how the fairness lines could be used to quantitatively analyze fair income distribution by comparing the scatter plots of the Cartesian coordinates of 75 countries in the sample where the abscissa is the income Gini index and the ordinate is the income shares by quintile ($Q_i^I$, i = 1, 2, …, 5) to the corresponding fairness lines in each quintile.



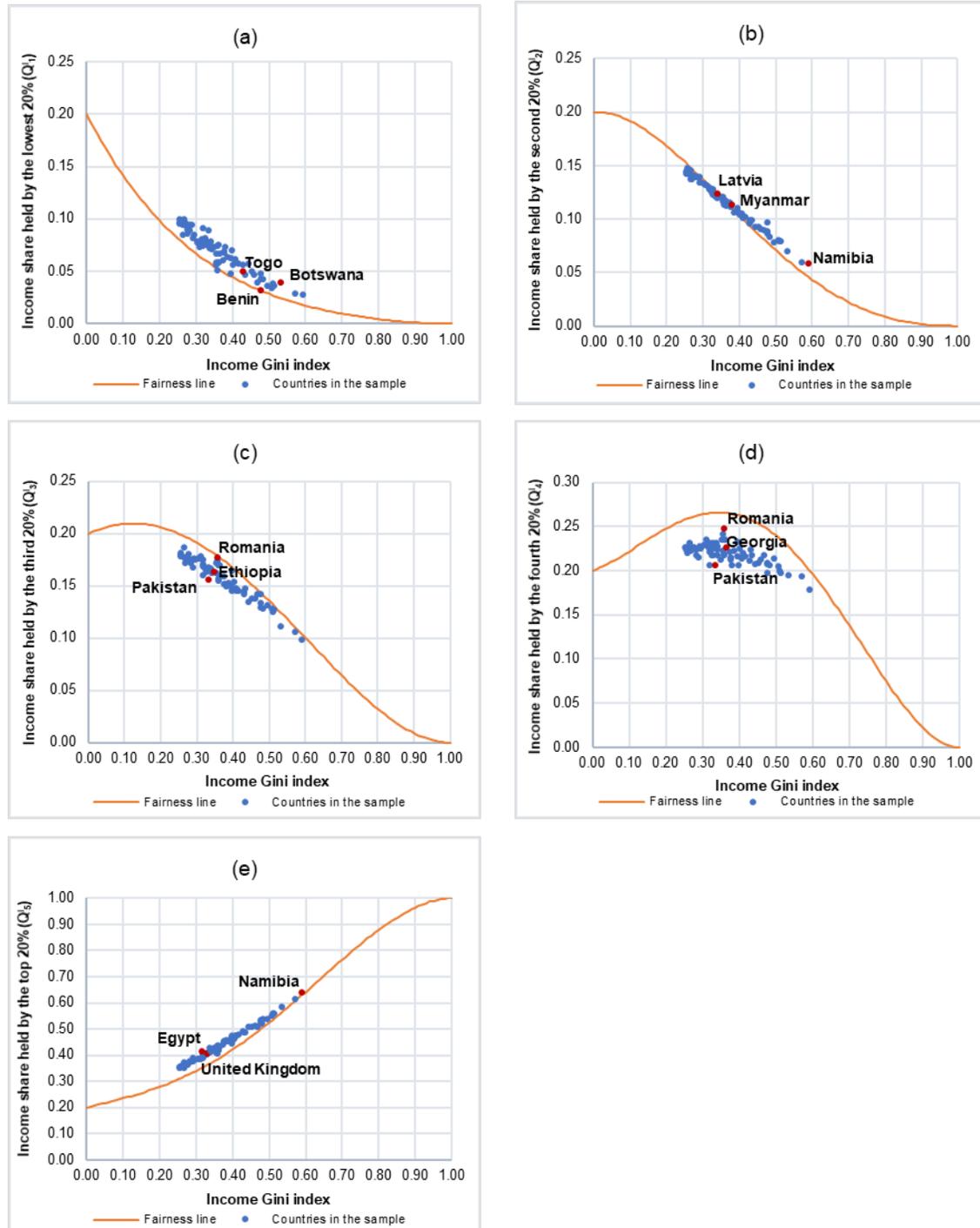

**Figure 6. Scatter plots of the Cartesian coordinates of the income Gini index and the income shares by quintile ($Q^I_i$, i = 1, 2, …, 5) of 75 countries. The fairness lines are used as a benchmark for analyzing whether the income share of each of 75 countries in each quintile is a fair share or not for a given value of the income Gini index of that country.** (a) Income share held by the lowest 20% ($Q^I_1$). (b) Income share held by the second 20% ($Q^I_2$). (c) Income share held by the third 20% ($Q^I_3$). (d) Income share held by the fourth 20% ($Q^I_4$). (e) Income share held by the top 20% ($Q^I_5$).



The overall results shown in Figure 6 indicate that the ordered-pairs (income Gini index, $Q^I_i$) of all 75 countries lie either above or below the fairness lines in all 5 quintiles, with those in the 1$^{st}$ and the 5$^{th}$ quintile being above their corresponding fairness lines (except Romania whose point is just below the fairness line in the 1$^{st}$ quintile) and those in the 3$^{rd}$ and the 4$^{th}$ quintile being below their corresponding fairness lines. For the 2$^{nd}$ quintile, there are 35 countries whose ordered-pairs (income Gini index, $Q^I_i$) lie above the fairness line while there are 40 countries whose ordered-pairs (income Gini index, $Q^I_i$) lie below the fairness line. The results from the sample of 75 countries suggest that, relative to our professional athlete's salary benchmark, the income earners in the 1$^{st}$ (except Romania), the 2$^{nd}$ (35 countries), and the 5$^{th}$ quintile receive income shares higher than the fair shares, with those in the 1$^{st}$ quintile receive the highest share as measured by the absolute values of the maximum (60.19%), the mean (29.98%), and the median (28.75%) of country's income share, whereas the income earners in the 2$^{nd}$ (40 countries), the 3$^{rd}$, and the 4$^{th}$ quintile receive income shares lower than the fair shares, with those in the 4$^{th}$ quintile receive the lowest share as measured by the absolute values of the minimum (22.80%), the mean (15.11%), and the median (14.78%) of country's income share.

The results from each quintile, starting from the 1$^{st}$ quintile where the income share is held by the lowest 20% (Figure 6a), show that, for an existing value of a country's Gini index, the income earners in 74 out of 75 countries have income shares higher than the fair shares as suggested by our professional athlete's salary benchmark. The country whose income share deviates from the fair share the most is Botswana (60.19%) while that of Benin deviates the least (1.58%), with the median country being Togo (28.75%). For the 2$^{nd}$ quintile where the income share is held by the second 20% (Figure 6b), Namibia is regarded as having the most unfair income share since the income earners in Namibia receive 26.97% higher than the fair share according to



our professional athlete's salary benchmark, whereas the least unfair income share country is Latvia, with merely 0.02% lower than the fair share. The median country in the $2^{nd}$ quintile is Myanmar whose income share is higher than the fair share by 2.96%. The results from the $3^{rd}$ quintile where the income share is held by the third 20% (Figure 6c) show that, compared to the professional athlete's salary benchmark, the income earners in Pakistan receive income share lower than the fair share the most (15.22%) while those in Romania receive income share lower than the fair share the least (0.68%), with Ethiopia being the median country whose income earners receive 9.67% lower than the fair share. For the $4^{th}$ quintile where the income share is held by the fourth 20% (Figure 6d), our results show that the income earners in Pakistan receive income share lower than the fair share the most as indicated by our professional athlete's salary benchmark (22.80%). The country whose income earners receive income share lower than the fair share the least is Romania (6.93%). Georgia is the median country in the $4^{th}$ quintile whose income earners receive income share lower than the fair share by 14.78%. Lastly, our results for the $5^{th}$ quintile where the income share is held by the top 20% (Figure 6e) indicate that, relative to our professional athlete's salary benchmark, the country whose income earners receive income share higher than the fair share the most is Egypt (17.29%) while the income earners in Namibia receive income share higher than the fair share the least (0.87%), with the United Kingdom being the median country whose income earners receive income share higher than the fair share by 11.37%. The actual income shares by quintile vs. the fair income shares by quintile as well as the values of the Gini index for all 75 countries in the sample are reported in Table A1 in the Appendix.

For the analysis of fair income distribution by quintile among different groups within a country and across countries, we choose Denmark and the Netherlands, on the criteria that both countries have the same Gini index (0.282) but differ in the income shares by quintile, in order to



demonstrate that our method not only would be able to identify which quintile the income share of a country is fair or not fair, but also suggest the level of fair income shares by quintile for that country. In addition, the income distribution by quintile between these two countries could be compared since they are gauged using the same professional athlete's salary benchmark. Our results are shown in Figure 7.



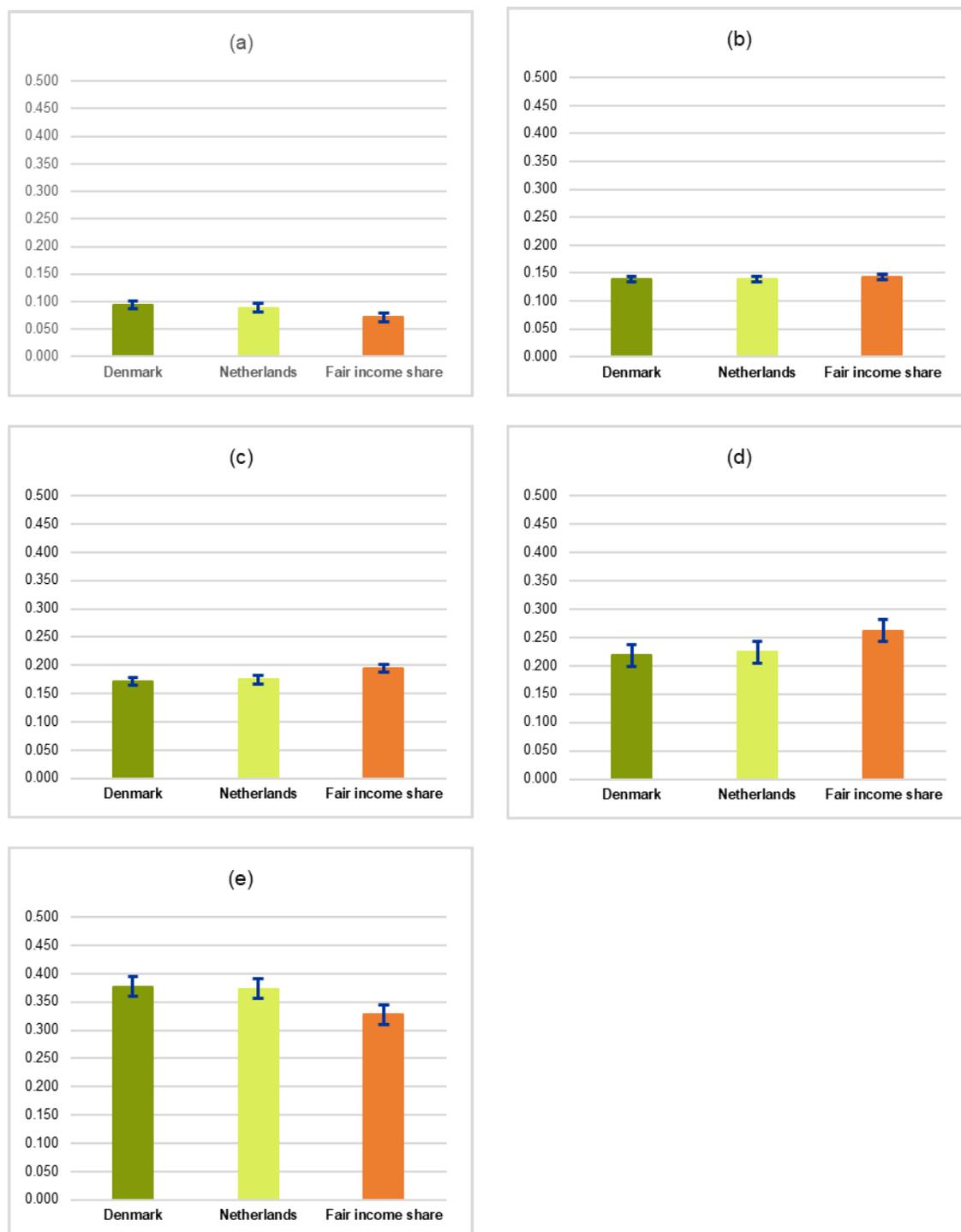

**Figure 7. Actual income shares by quintile of Denmark and the Netherlands vs. fair income shares by quintile based on the professional athlete's salary benchmark.** (a) The lowest 20%. (b) The second 20%. (c) The third 20%. (d) The fourth (20%). (e) The top 20%. Note that the Gini index of Denmark and the Netherlands is equal to 0.282. The standard errors are indicated by the bars.



We can see from Figure 7 that both Denmark and the Netherlands share a common pattern of income shares by quintile (in decimals), only slightly differ in the degree of fairness. Given the existing value of the Gini index which is equal to 0.282 in both countries, our professional athlete's salary benchmark indicates that the fair income share for the lowest 20% should be equal to 0.072 (Figure 7a). However, the actual income shares in the 1$^{st}$ quintile of Denmark and the Netherlands are equal to 0.094 and 0.089, respectively. The fair income share for the second 20% should be 0.143 according to our professional athlete' salary benchmark, but both Denmark and the Netherlands have slightly lower income shares in this quintile at 0.139 (Figure 7b). For the third 20%, the actual income share in Denmark equals 0.172 and that in the Netherlands equals 0.175 which are lower than the fair income share based on our professional athlete's salary benchmark which suggests that it should equal 0.195 (Figure 7c). While our professional athlete's salary benchmark indicates that the fair income share for the fourth 40% should equal 0.262, both Denmark and the Netherlands have lower actual income shares at 0.218 and 0.224, respectively (Figure 7d). For the top 20%, our professional athlete's salary benchmark suggests that the fair income share should be 0.328 (Figure 7e). However, both Denmark and the Netherlands have higher levels of actual income shares at 0.377 and 0.373, respectively.

## 4. Discussion

Given that rising income inequality is a persistent cause of concern, the greater focus is needed to lower income inequality (the United Nations, 2019). While the SDG 10 (Target 10.1) calls for actions to reduce income inequality within and among countries by progressively achieving and sustaining income growth of the bottom 40% of the population at a rate higher than the national average by 2030, it should be aware that a low level of income inequality could



potentially have adverse effects on economic development (Grigoli & Robles, 2017). Imagine a society where everyone has more or less similar income. In such a society, there should be no incentive for anyone to be creative or try to do things differently because no matter how hard they try or what they do and/or invent, there would be no extra benefits. This could put that society into an incentive trap which has negative impacts on productivity and economic growth (Sitthiyot & Holasut, 2016). In addition, contrary to the view that high income inequality is an ethical and moral concern across cultures around the world (the United Nations, 2020b), a wealth of empirical research in behavioral economics and psychology as discussed in this study has shown that people are not troubled by income inequality per se. Indeed, they prefer fair income distributions, not equal ones, both in laboratory conditions and in the real world (Starmans et al., 2017). People even support substantial income inequalities if they perceive as fair (Trump, 2020) Thus, from both theoretical and practical points of view, it is very important to know what level of inequality in income distribution is considered as fair.

In this study, we devise an alternative benchmark for measuring whether the existing level of income distribution of a country is fair or not, and if not, what fair income distribution in that country should be. Our benchmark is constructed based on the concepts of procedural justice, distributive justice, and power of authority in the realm of 11 well-known professional sports where the allocations of athlete's salary are the outcomes of fair rules, individual and/or team performance, and luck. While the political economy of what levels of income differences countries would accept as fair remains an unexplored territory (Rogoff, 2012), with the exception of a few studies by Venkatasubramanian (2009; 2010; 2019) and Venkatasubramanian et al. (2015) who propose the use of entropy in statistical mechanics and information theory as a quantitative measure of fairness in income distribution and a study by Park & Kim (2021) who



introduce the concept of a feasible income equality that maximizes the total social welfare and show that an optimal income distribution representing the feasible equality could be modeled by using the sigmoid welfare function and the Boltzmann income distribution, to our knowledge, no study has attempted to quantitatively measure fair income distribution across countries using the benchmark that is derived from the statistical relationship between the salary shares of the professional athlete in each quintile and the Gini index for the professional athlete's salary in this way before. Our benchmark also satisfies the no-envy principle of fair allocation introduced by Tinbergen (1930) and the general consensus or the international norm criterion of a meaningful benchmark according to the OHCHR (2012). At the very least, the benchmark and the findings presented in our study could be viewed as another proof of concept that fair income distribution could be quantified. If future research could find other types of data that are fairer based on the concepts of procedural justice, distributive justice, and power of authority, and more universally accepted than the data on the distributions of athlete's salary in professional sports employed in this study, our method could still be used to develop the fairness benchmarks, provided such data can be numerically ranked. This is because we can employ an appropriate functional form for the Lorenz curve to fit those data, calculate their respective values of the Gini index and the quantile shares, and find the fairness benchmarks by estimating the statistical relationships between the values of the Gini index and the same quantile shares across those data.

To demonstrate how our benchmark works, we use the data on the Gini index and the income shares by quintile in 2015 from the World Bank (2020) comprising 75 countries. The overall results indicate that the income earners in the lowest 20%, the second 20% (35 countries), and the top 20% in countries in the sample (except those in the lowest 20% in Romania) receive income shares higher than the fair shares according to our professional athlete's salary



benchmark, with those in the lowest 20% receive the highest share as measured by the maximum, the mean, and the median values of country's income share. For those in the second 20% (40 countries), the third 20% and the fourth 20%, our results suggest that, the income earners in countries in the sample receive income shares lower than the fair shares based on the professional athlete's salary benchmark, with those in the fourth 20% receive the lowest share as measured by the minimum, the mean, and the median values of country's income share. Using Denmark and the Netherlands as examples, we demonstrate that our method could be used to compare fair income distribution among different groups within a country and across countries, provided that they share the same Gini index.

Furthermore, our professional athlete's salary benchmark could be employed as a guideline for setting targets for fair income distribution as measured by the Gini index and the income shares by quintile. For example, supposing that an acceptable value of the Gini index is 0.226 and the minimum level of income share for the bottom 40% must be at least 25% of the total income, our benchmark suggests that the fair income shares by quintile should be such that the top 20% and the fourth 20% should receive income shares around 29.4% and 25.3%, respectively. The third 20% would receive income share about 20.3% whereas the second 20% and the lowest 20% should receive income shares roughly 16.1% and 8.9%, respectively. Based on this example, the bottom 40% would receive income share around 25% and the income gap between the top 20% and the bottom 20% would be approximately 3.29 times. In the real world, there are infinite combinations of income shares by quintile that could yield the same value of the Gini index but there is only one of them that is regarded as fair. Thus, by using our benchmark, different countries could choose different combinations of the Gini index and the fair



income shares by quintile that are appropriate for their own context before formulating and conducting policies in order to achieve the SDG 10 and other related SDGs.

While one of the major concerns about income inequality has been on the income earners whose income shares are in the bottom 40% (Keeley, 2015; the United Nations, 2015; the World Bank, 2015; the Organization for Economic Co-operation and Development (OECD), 2019; the United Nations, 2020a; the United Nations, 2020b; the World Bank, 2020b), our empirical findings from 75 countries indicate that, *on the fairness ground*, the majority of the income earners in the bottom 40% is in a relatively better position, as measured by the income shares they receive, than those in the third 20% and the fourth 20% who receive income shares lower than the fair shares. These results are well supported by the stylized facts known as "the Elephant Chart" published in Alvaredo et al. (2017, p. 13) and "the Loch Ness Monster Chart" published in Kharas & Seidel (2018, p. 12), both of which represent the relationship between the income growth and the quantile global income distribution. Both "the Elephant Chart" and "the Loch Ness Monster Chart", according to these 2 studies, clearly illustrate that the income growth of the income earners in the third 20% and the fourth 20% is lower than that of the income earners in other 3 quintiles. Another study by the OECD (2019) also finds similar results in that the middle-income households in the OECD countries have experienced dismal income growth or even stagnation in some countries. According to the OECD (2019), this has fueled perceptions that the current socio-economic system is *unfair* [emphasis in original] and that the middle class has not benefited from economic growth in proportion to its contribution. These empirical findings indicate that, in addition to the existing tools or diagrams used to study issues in income distribution, fairness and income inequality could be analyzed simultaneously through the lens of procedural justice, distributive justice, and power of authority using the fairness benchmark that



is derived from the relationship between the salary shares of the professional athlete by quintile and the Gini index for the professional athlete's salary.

Despite the empirical findings with regard to the level of fair income share of the bottom 40%, we would like to emphasize that this by no means implies that we should not continue to pursue policies in order to increase and sustain the income growth of the bottom 40% as stated in the Target 10.1 of the SDG 10. Reducing income inequality is found to be positively correlated with a number of socio-economic indicators that are fundamental for sustainable development Wilkinson & Pickett (2007). However, as noted by Frankfurt (2015) and Starmans et al. (2017), worries about income inequality are often conflated with worries about poverty and unfairness whereas these concerns should be thought of as distinct. According to Starmans et al. (2017), if income inequality in itself is not really what is bothering people as supported by laboratory and real-world studies, then we would be better off by clearly distinguishing these concerns, and shifting our attention to the problem that matters to us more which, in this case, is income inequity. After all, from the viewpoint of morality as argued by Frankfurt (2015, p. 7), it is not important that everyone should have *the same* [emphasis in original]. What is morally important is that everyone has *enough* [emphasis in original], provided that the income a person has results from fair processes and is proportional to his/her contributions. According to Frankfurt (2015, p. 7), if everyone has enough income, it should be of no special or deliberate concern whether some people have more income than others.

From our viewpoint, the grand challenge for everyone involved in formulating and implementing policies in order to achieve fair income distribution societies, the SGD 10, and other SDGs that are directly or indirectly related to the SDG 10 is not only to make processes with regard to the income allocations and the distributions of income fair but also to make the



authorities being transparent and trustworthy when it comes to the issues of enacting rules, resolving disputes, and allocating resources. Moreover, people have to be educated and nudged so that they understand what make such processes and allocations fair. According to Dunham et al. (2018), this requires an understanding that a process can be fair despite producing an unequal outcome. It also requires understanding the concept of randomness since some processes can be fair by the virtue of being unpredictable and thereby not exploitable by involved parties. Furthermore, people themselves must be able to override their own selfishness and come to prefer processes that are procedurally fair and allocations that are distributively fair, even when these processes and allocations might materially put them in disadvantage positions. This is a tall order task. What forms of governance and institution that are capable of pulling this off and how long it would take remain unclear. Last but not least, we hope that our benchmark could be used as an alternative tool for analyzing fair size distributions of resources and other non-negative quantities in other scientific disciplines.


**Author Contributions:** Conceptualization: TS; Data curation: TS; Methodology: TS, KH; Formal analysis: TS; Validation: KH; Writing – original draft: TS; Writing – review and editing: TS, KH.

**Funding:** This research received no external funding.

**Data Availability Statement:** All data generated and/or analyzed during this study are included in this article and can be accessed from the data repository and the World Bank websites as shown in the References.

**Acknowledgments:** TS is grateful to Dr. Suradit Holasut for guidance and comments.

**Conflicts of Interest:** The authors declare no conflict of interest.


# Appendix

**Table A1. Actual income shares by quintile vs. fair income shares by quintile and the Gini index of 75 countries**.

| Country | Gini index | Income share | Top 20% | %Δ | Fourth 20% | %Δ | Third 20% | %Δ | Second 20% | %Δ | Lowest 20% | %Δ |
|---|---|---|---|---|---|---|---|---|---|---|---|---|
| 1. Slovenia | 0.254 | Actual | 0.351 | 13.18 | 0.226 | -12.39 | 0.182 | -8.80 | 0.145 | -4.72 | 0.096 | 19.77 |
| | | Fair | 0.310 | | 0.258 | | 0.200 | | 0.152 | | 0.080 | |
| 2. Ukraine | 0.255 | Actual | 0.356 | 14.57 | 0.224 | -13.22 | 0.178 | -10.74 | 0.143 | -5.84 | 0.100 | 25.24 |
| | | Fair | 0.311 | | 0.258 | | 0.199 | | 0.152 | | 0.080 | |
| 3. Belarus | 0.256 | Actual | 0.355 | 14.02 | 0.226 | -12.50 | 0.180 | -9.66 | 0.143 | -5.65 | 0.097 | 21.95 |
| | | Fair | 0.311 | | 0.258 | | 0.199 | | 0.152 | | 0.080 | |
| 4. Czech Republic | 0.259 | Actual | 0.359 | 14.61 | 0.219 | -15.37 | 0.178 | -10.46 | 0.147 | -2.39 | 0.097 | 23.38 |
| | | Fair | 0.313 | | 0.259 | | 0.199 | | 0.151 | | 0.079 | |
| 5. Slovak Republic | 0.265 | Actual | 0.350 | 10.40 | 0.232 | -10.65 | 0.187 | -5.47 | 0.146 | -1.80 | 0.085 | 10.67 |
| | | Fair | 0.317 | | 0.260 | | 0.198 | | 0.149 | | 0.077 | |
| 6. Kosovo | 0.265 | Actual | 0.361 | 13.87 | 0.227 | -12.58 | 0.177 | -10.53 | 0.140 | -5.84 | 0.095 | 23.69 |
| | | Fair | 0.317 | | 0.260 | | 0.198 | | 0.149 | | 0.077 | |
| 7. Kazakhstan | 0.268 | Actual | 0.369 | 15.69 | 0.222 | -14.64 | 0.172 | -12.84 | 0.137 | -7.25 | 0.100 | 31.73 |
| | | Fair | 0.319 | | 0.260 | | 0.197 | | 0.148 | | 0.076 | |
| 8. Moldova | 0.270 | Actual | 0.369 | 15.22 | 0.223 | -14.35 | 0.175 | -11.17 | 0.137 | -6.84 | 0.097 | 28.77 |
| | | Fair | 0.320 | | 0.260 | | 0.197 | | 0.147 | | 0.075 | |
| 9. Finland | 0.271 | Actual | 0.367 | 14.37 | 0.224 | -14.01 | 0.175 | -11.09 | 0.140 | -4.59 | 0.094 | 25.28 |
| | | Fair | 0.321 | | 0.260 | | 0.197 | | 0.147 | | 0.075 | |

Note: The Gini index and income shares are shown in decimals.

*(Continued).*



**Table A1. Actual income shares by quintile vs. fair income shares by quintile and the Gini index of 75 countries**. (Continued).

| Country | Gini index | Income share | Top 20% | %Δ | Fourth 20% | %Δ | Third 20% | %Δ | Second 20% | %Δ | Lowest 20% | %Δ |
|---|---|---|---|---|---|---|---|---|---|---|---|---|
| 10. Norway | 0.275 | Actual | 0.365 | 12.82 | 0.227 | -13.03 | 0.177 | -9.76 | 0.141 | -3.06 | 0.090 | 21.84 |
|  |  | Fair | 0.324 |  | 0.261 |  | 0.196 |  | 0.145 |  | 0.074 |  |
| 11. Belgium | 0.277 | Actual | 0.365 | 12.36 | 0.227 | -13.12 | 0.181 | -7.56 | 0.140 | -3.31 | 0.086 | 17.34 |
|  |  | Fair | 0.325 |  | 0.261 |  | 0.196 |  | 0.145 |  | 0.073 |  |
| 12. Denmark | 0.282 | Actual | 0.377 | 14.87 | 0.218 | -16.76 | 0.172 | -11.75 | 0.139 | -2.90 | 0.094 | 30.80 |
|  |  | Fair | 0.328 |  | 0.262 |  | 0.195 |  | 0.143 |  | 0.072 |  |
| 13. Netherlands | 0.282 | Actual | 0.373 | 13.65 | 0.224 | -14.46 | 0.175 | -10.21 | 0.139 | -2.90 | 0.089 | 23.84 |
|  |  | Fair | 0.328 |  | 0.262 |  | 0.195 |  | 0.143 |  | 0.072 |  |
| 14. Kyrgyz Republic | 0.290 | Actual | 0.388 | 16.29 | 0.215 | -18.18 | 0.168 | -13.15 | 0.134 | -4.64 | 0.095 | 36.42 |
|  |  | Fair | 0.334 |  | 0.263 |  | 0.193 |  | 0.141 |  | 0.070 |  |
| 15. Sweden | 0.292 | Actual | 0.376 | 12.23 | 0.228 | -13.30 | 0.176 | -8.83 | 0.139 | -0.61 | 0.082 | 18.69 |
|  |  | Fair | 0.335 |  | 0.263 |  | 0.193 |  | 0.140 |  | 0.069 |  |
| 16. Malta | 0.294 | Actual | 0.381 | 13.25 | 0.225 | -14.50 | 0.175 | -9.17 | 0.134 | -3.73 | 0.085 | 24.01 |
|  |  | Fair | 0.336 |  | 0.263 |  | 0.193 |  | 0.139 |  | 0.069 |  |
| 17. Hungary | 0.304 | Actual | 0.384 | 11.78 | 0.229 | -13.27 | 0.176 | -7.71 | 0.133 | -2.11 | 0.078 | 18.41 |
|  |  | Fair | 0.344 |  | 0.264 |  | 0.191 |  | 0.136 |  | 0.066 |  |
| 18. Austria | 0.305 | Actual | 0.384 | 11.54 | 0.229 | -13.29 | 0.176 | -7.61 | 0.132 | -2.60 | 0.079 | 20.41 |
|  |  | Fair | 0.344 |  | 0.264 |  | 0.191 |  | 0.136 |  | 0.066 |  |

Note: The Gini index and income shares are shown in decimals.

*(Continued).*



**Table A1. Actual income shares by quintile vs. fair income shares by quintile and the Gini index of 75 countries**. (Continued).

| Country | Gini index | Income share | Top 20% | %Δ | Fourth 20% | %Δ | Third 20% | %Δ | Second 20% | %Δ | Lowest 20% | %Δ |
|---|---|---|---|---|---|---|---|---|---|---|---|---|
| 19 Croatia | 0.311 | Actual | 0.384 | 10.14 | 0.235 | -11.16 | 0.178 | -5.95 | 0.131 | -1.88 | 0.073 | 13.96 |
| | | Fair | 0.349 | | 0.265 | | 0.189 | | 0.134 | | 0.064 | |
| 20. Germany | 0.317 | Actual | 0.397 | 12.44 | 0.226 | -14.68 | 0.170 | -9.57 | 0.129 | -1.90 | 0.078 | 24.73 |
| | | Fair | 0.353 | | 0.265 | | 0.188 | | 0.131 | | 0.063 | |
| 21. Egypt, Arab Rep. | 0.318 | Actual | 0.415 | 17.29 | 0.206 | -22.25 | 0.160 | -14.80 | 0.128 | -2.41 | 0.091 | 46.10 |
| | | Fair | 0.354 | | 0.265 | | 0.188 | | 0.131 | | 0.062 | |
| 22. Ireland | 0.318 | Actual | 0.402 | 13.61 | 0.221 | -16.58 | 0.169 | -10.00 | 0.129 | -1.64 | 0.080 | 28.44 |
| | | Fair | 0.354 | | 0.265 | | 0.188 | | 0.131 | | 0.062 | |
| 23. Poland | 0.318 | Actual | 0.392 | 10.78 | 0.231 | -12.81 | 0.175 | -6.81 | 0.128 | -2.41 | 0.074 | 18.81 |
| | | Fair | 0.354 | | 0.265 | | 0.188 | | 0.131 | | 0.062 | |
| 24. Switzerland | 0.323 | Actual | 0.402 | 12.41 | 0.226 | -14.77 | 0.168 | -10.01 | 0.125 | -3.45 | 0.078 | 27.78 |
| | | Fair | 0.358 | | 0.265 | | 0.187 | | 0.129 | | 0.061 | |
| 25. Armenia | 0.324 | Actual | 0.407 | 13.57 | 0.218 | -17.80 | 0.167 | -10.44 | 0.126 | -2.42 | 0.082 | 34.88 |
| | | Fair | 0.358 | | 0.265 | | 0.186 | | 0.129 | | 0.061 | |
| 26. Estonia | 0.327 | Actual | 0.404 | 12.01 | 0.232 | -12.56 | 0.164 | -11.74 | 0.125 | -2.43 | 0.075 | 24.87 |
| | | Fair | 0.361 | | 0.265 | | 0.186 | | 0.128 | | 0.060 | |
| 27. France | 0.327 | Actual | 0.409 | 13.40 | 0.217 | -18.21 | 0.167 | -10.12 | 0.128 | -0.09 | 0.079 | 31.53 |
| | | Fair | 0.361 | | 0.265 | | 0.186 | | 0.128 | | 0.060 | |

Note: The Gini index and income shares are shown in decimals.

*(Continued).*



**Table A1. Actual income shares by quintile vs. fair income shares by quintile and the Gini index of 75 countries**. (Continued).

| Country | Gini index | Income share | Top 20% | %Δ | Fourth 20% | %Δ | Third 20% | %Δ | Second 20% | %Δ | Lowest 20% | %Δ |
|---|---|---|---|---|---|---|---|---|---|---|---|---|
| 28. Tunisia | 0.328 | Actual | 0.409 | 13.15 | 0.225 | -15.21 | 0.165 | -11.09 | 0.123 | -3.74 | 0.078 | 30.39 |
| | | Fair | 0.361 | | 0.265 | | 0.186 | | 0.128 | | 0.060 | |
| 29. United Kingdom | 0.332 | Actual | 0.406 | 11.37 | 0.230 | -13.36 | 0.168 | -9.03 | 0.122 | -3.49 | 0.075 | 27.42 |
| | | Fair | 0.365 | | 0.265 | | 0.185 | | 0.126 | | 0.059 | |
| 30. Pakistan | 0.335 | Actual | 0.428 | 16.65 | 0.205 | -22.80 | 0.156 | -15.22 | 0.122 | -2.71 | 0.089 | 53.06 |
| | | Fair | 0.367 | | 0.266 | | 0.184 | | 0.125 | | 0.058 | |
| 31. Luxembourg | 0.338 | Actual | 0.410 | 11.02 | 0.231 | -13.02 | 0.167 | -8.89 | 0.121 | -2.72 | 0.072 | 25.34 |
| | | Fair | 0.369 | | 0.266 | | 0.183 | | 0.124 | | 0.057 | |
| 32. Cyprus | 0.340 | Actual | 0.421 | 13.51 | 0.217 | -18.30 | 0.162 | -11.39 | 0.121 | -2.18 | 0.079 | 38.65 |
| | | Fair | 0.371 | | 0.266 | | 0.183 | | 0.124 | | 0.057 | |
| 33. Tajikistan | 0.340 | Actual | 0.417 | 12.43 | 0.224 | -15.66 | 0.164 | -10.30 | 0.120 | -2.99 | 0.074 | 29.88 |
| | | Fair | 0.371 | | 0.266 | | 0.183 | | 0.124 | | 0.057 | |
| 34. Latvia | 0.342 | Actual | 0.415 | 11.41 | 0.228 | -14.16 | 0.164 | -10.07 | 0.123 | -0.02 | 0.071 | 25.64 |
| | | Fair | 0.372 | | 0.266 | | 0.182 | | 0.123 | | 0.057 | |
| 35. Ethiopia | 0.350 | Actual | 0.430 | 13.46 | 0.212 | -20.18 | 0.163 | -9.67 | 0.121 | 0.58 | 0.073 | 33.48 |
| | | Fair | 0.379 | | 0.266 | | 0.180 | | 0.120 | | 0.055 | |
| 36. Italy | 0.354 | Actual | 0.413 | 8.04 | 0.235 | -11.50 | 0.172 | -4.16 | 0.121 | 1.74 | 0.059 | 9.67 |
| | | Fair | 0.382 | | 0.266 | | 0.179 | | 0.119 | | 0.054 | |

Note: The Gini index and income shares are shown in decimals.

*(Continued).*



**Table A1. Actual income shares by quintile vs. fair income shares by quintile and the Gini index of 75 countries**. (Continued).

| Country | Gini index | Income share | Top 20% | %Δ | Fourth 20% | %Δ | Third 20% | %Δ | Second 20% | %Δ | Lowest 20% | %Δ |
|---|---|---|---|---|---|---|---|---|---|---|---|---|
| 37. Portugal | 0.355 | Actual | 0.427 | 11.46 | 0.223 | -16.01 | 0.163 | -9.05 | 0.120 | 1.19 | 0.067 | 25.05 |
| | | Fair | 0.383 | | 0.266 | | 0.179 | | 0.119 | | 0.054 | |
| 38. North Macedonia | 0.356 | Actual | 0.411 | 7.05 | 0.242 | -8.85 | 0.174 | -2.77 | 0.117 | -1.06 | 0.056 | 4.95 |
| | | Fair | 0.384 | | 0.265 | | 0.179 | | 0.118 | | 0.053 | |
| 39. Gambia, The | 0.359 | Actual | 0.436 | 12.82 | 0.218 | -17.86 | 0.157 | -11.90 | 0.116 | -1.05 | 0.074 | 40.41 |
| | | Fair | 0.386 | | 0.265 | | 0.178 | | 0.117 | | 0.053 | |
| 40. Romania | 0.359 | Actual | 0.407 | 5.32 | 0.247 | -6.93 | 0.177 | -0.68 | 0.118 | 0.66 | 0.051 | -3.23 |
| | | Fair | 0.386 | | 0.265 | | 0.178 | | 0.117 | | 0.053 | |
| 41. Greece | 0.360 | Actual | 0.418 | 7.93 | 0.235 | -11.44 | 0.170 | -4.48 | 0.118 | 0.95 | 0.059 | 12.41 |
| | | Fair | 0.387 | | 0.265 | | 0.178 | | 0.117 | | 0.052 | |
| 42. Thailand | 0.360 | Actual | 0.438 | 13.09 | 0.219 | -17.47 | 0.155 | -12.90 | 0.113 | -3.33 | 0.075 | 42.90 |
| | | Fair | 0.387 | | 0.265 | | 0.178 | | 0.117 | | 0.052 | |
| 43. Spain | 0.362 | Actual | 0.421 | 8.23 | 0.235 | -11.42 | 0.170 | -4.20 | 0.117 | 0.68 | 0.058 | 11.43 |
| | | Fair | 0.389 | | 0.265 | | 0.177 | | 0.116 | | 0.052 | |
| 44. Georgia | 0.365 | Actual | 0.434 | 10.85 | 0.226 | -14.78 | 0.159 | -10.01 | 0.113 | -1.90 | 0.067 | 30.33 |
| | | Fair | 0.392 | | 0.265 | | 0.177 | | 0.115 | | 0.051 | |
| 45. Lithuania | 0.374 | Actual | 0.441 | 10.45 | 0.221 | -16.51 | 0.159 | -8.81 | 0.116 | 3.45 | 0.061 | 23.18 |
| | | Fair | 0.399 | | 0.265 | | 0.174 | | 0.112 | | 0.050 | |

Note: The Gini index and income shares are shown in decimals.

*(Continued)*.



**Table A1. Actual income shares by quintile vs. fair income shares by quintile and the Gini index of 75 countries**. (Continued).

| Country | Gini index | Income share | Top 20% | %Δ | Fourth 20% | %Δ | Third 20% | %Δ | Second 20% | %Δ | Lowest 20% | %Δ |
|---|---|---|---|---|---|---|---|---|---|---|---|---|
| 46. Tonga | 0.376 | Actual | 0.454 | 13.21 | 0.212 | -19.87 | 0.152 | -12.56 | 0.114 | 2.29 | 0.068 | 38.47 |
| | | Fair | 0.401 | | 0.265 | | 0.174 | | 0.111 | | 0.049 | |
| 47. Russian Federation | 0.377 | Actual | 0.453 | 12.71 | 0.215 | -18.72 | 0.152 | -12.43 | 0.111 | -0.10 | 0.069 | 41.10 |
| | | Fair | 0.402 | | 0.265 | | 0.174 | | 0.111 | | 0.049 | |
| 48. Myanmar | 0.381 | Actual | 0.457 | 12.72 | 0.206 | -22.04 | 0.150 | -13.04 | 0.113 | 2.96 | 0.073 | 51.80 |
| | | Fair | 0.405 | | 0.264 | | 0.172 | | 0.110 | | 0.048 | |
| 49. China | 0.386 | Actual | 0.454 | 10.76 | 0.223 | -15.47 | 0.153 | -10.60 | 0.106 | -1.90 | 0.064 | 35.91 |
| | | Fair | 0.410 | | 0.264 | | 0.171 | | 0.108 | | 0.047 | |
| 50. Iran, Islamic Rep. | 0.395 | Actual | 0.464 | 10.99 | 0.216 | -17.85 | 0.150 | -11.07 | 0.107 | 1.90 | 0.063 | 38.97 |
| | | Fair | 0.418 | | 0.263 | | 0.169 | | 0.105 | | 0.045 | |
| 51. Serbia | 0.396 | Actual | 0.444 | 5.97 | 0.233 | -11.35 | 0.165 | -2.01 | 0.110 | 5.09 | 0.048 | 6.33 |
| | | Fair | 0.419 | | 0.263 | | 0.168 | | 0.105 | | 0.045 | |
| 52. Indonesia | 0.397 | Actual | 0.473 | 12.65 | 0.206 | -21.59 | 0.146 | -13.15 | 0.106 | 1.60 | 0.070 | 55.73 |
| | | Fair | 0.420 | | 0.263 | | 0.168 | | 0.104 | | 0.045 | |
| 53. Uruguay | 0.402 | Actual | 0.461 | 8.59 | 0.225 | -14.17 | 0.154 | -7.61 | 0.104 | 1.32 | 0.056 | 27.26 |
| | | Fair | 0.425 | | 0.262 | | 0.167 | | 0.103 | | 0.044 | |
| 54. El Salvador | 0.406 | Actual | 0.472 | 10.21 | 0.214 | -18.20 | 0.148 | -10.60 | 0.105 | 3.65 | 0.061 | 41.01 |
| | | Fair | 0.428 | | 0.262 | | 0.166 | | 0.101 | | 0.043 | |

Note: The Gini index and income shares are shown in decimals.

*(Continued).*



**Table A1. Actual income shares by quintile vs. fair income shares by quintile and the Gini index of 75 countries**. (Continued).

| Country | Gini index | Income share | Top 20% | %Δ | Fourth 20% | %Δ | Third 20% | %Δ | Second 20% | %Δ | Lowest 20% | %Δ |
|---|---|---|---|---|---|---|---|---|---|---|---|---|
| 55. Kenya | 0.408 | Actual | 0.475 | 10.43 | 0.215 | -17.74 | 0.146 | -11.50 | 0.103 | 2.35 | 0.062 | 44.55 |
| | | Fair | 0.430 | | 0.261 | | 0.165 | | 0.101 | | 0.043 | |
| 56. Malaysia | 0.410 | Actual | 0.473 | 9.48 | 0.220 | -15.74 | 0.148 | -9.97 | 0.101 | 1.04 | 0.058 | 36.39 |
| | | Fair | 0.432 | | 0.261 | | 0.164 | | 0.100 | | 0.043 | |
| 57. Cote d'Ivoire | 0.415 | Actual | 0.478 | 9.43 | 0.216 | -17.04 | 0.146 | -10.40 | 0.102 | 3.77 | 0.057 | 36.95 |
| | | Fair | 0.437 | | 0.260 | | 0.163 | | 0.098 | | 0.042 | |
| 58. Turkey | 0.429 | Actual | 0.492 | 9.24 | 0.211 | -18.22 | 0.142 | -10.56 | 0.099 | 5.73 | 0.056 | 42.95 |
| | | Fair | 0.450 | | 0.258 | | 0.159 | | 0.094 | | 0.039 | |
| 59. Togo | 0.431 | Actual | 0.486 | 7.43 | 0.224 | -13.06 | 0.144 | -8.95 | 0.095 | 2.17 | 0.050 | 28.75 |
| | | Fair | 0.452 | | 0.258 | | 0.158 | | 0.093 | | 0.039 | |
| 60. Peru | 0.434 | Actual | 0.487 | 6.95 | 0.218 | -15.20 | 0.148 | -5.88 | 0.099 | 7.62 | 0.047 | 22.62 |
| | | Fair | 0.455 | | 0.257 | | 0.157 | | 0.092 | | 0.038 | |
| 61. Philippines | 0.444 | Actual | 0.509 | 9.37 | 0.207 | -18.84 | 0.135 | -12.43 | 0.093 | 4.83 | 0.057 | 55.39 |
| | | Fair | 0.465 | | 0.255 | | 0.154 | | 0.089 | | 0.037 | |
| 62. Dominican Republic | 0.452 | Actual | 0.511 | 7.90 | 0.209 | -17.48 | 0.138 | -9.00 | 0.093 | 7.99 | 0.050 | 41.22 |
| | | Fair | 0.474 | | 0.253 | | 0.152 | | 0.086 | | 0.035 | |
| 63. Ecuador | 0.460 | Actual | 0.513 | 6.46 | 0.213 | -15.25 | 0.138 | -7.45 | 0.091 | 8.93 | 0.046 | 34.63 |
| | | Fair | 0.482 | | 0.251 | | 0.149 | | 0.084 | | 0.034 | |

Note: The Gini index and income shares are shown in decimals.

*(Continued).*



**Table A1. Actual income shares by quintile vs. fair income shares by quintile and the Gini index of 75 countries**. (Continued).

| Country | Gini index | Income share | Top 20% | %Δ | Fourth 20% | %Δ | Third 20% | %Δ | Second 20% | %Δ | Lowest 20% | %Δ |
|---|---|---|---|---|---|---|---|---|---|---|---|---|
| 64. Bolivia | 0.467 | Actual | 0.511 | 4.45 | 0.218 | -12.63 | 0.142 | -3.30 | 0.090 | 10.69 | 0.039 | 17.79 |
| | | Fair | 0.489 | | 0.250 | | 0.147 | | 0.081 | | 0.033 | |
| 65. Paraguay | 0.476 | Actual | 0.527 | 5.65 | 0.209 | -15.40 | 0.134 | -6.88 | 0.087 | 10.88 | 0.043 | 35.26 |
| | | Fair | 0.499 | | 0.247 | | 0.144 | | 0.078 | | 0.032 | |
| 66. Chile | 0.477 | Actual | 0.536 | 7.23 | 0.197 | -20.17 | 0.130 | -9.45 | 0.089 | 13.89 | 0.048 | 51.68 |
| | | Fair | 0.500 | | 0.247 | | 0.144 | | 0.078 | | 0.032 | |
| 67. Benin | 0.478 | Actual | 0.521 | 4.00 | 0.208 | -15.61 | 0.142 | -0.87 | 0.096 | 23.34 | 0.032 | 1.58 |
| | | Fair | 0.501 | | 0.246 | | 0.143 | | 0.078 | | 0.032 | |
| 68. Costa Rica | 0.484 | Actual | 0.539 | 6.22 | 0.206 | -15.82 | 0.128 | -9.38 | 0.084 | 10.58 | 0.043 | 40.29 |
| | | Fair | 0.507 | | 0.245 | | 0.141 | | 0.076 | | 0.031 | |
| 69. Honduras | 0.496 | Actual | 0.540 | 3.73 | 0.214 | -11.18 | 0.132 | -3.81 | 0.078 | 7.94 | 0.036 | 24.12 |
| | | Fair | 0.521 | | 0.241 | | 0.137 | | 0.072 | | 0.029 | |
| 70. Panama | 0.508 | Actual | 0.553 | 3.57 | 0.205 | -13.45 | 0.128 | -3.86 | 0.080 | 16.56 | 0.034 | 23.97 |
| | | Fair | 0.534 | | 0.237 | | 0.133 | | 0.069 | | 0.027 | |
| 71. Colombia | 0.511 | Actual | 0.559 | 4.03 | 0.200 | -15.18 | 0.125 | -5.38 | 0.079 | 16.63 | 0.038 | 40.53 |
| | | Fair | 0.537 | | 0.236 | | 0.132 | | 0.068 | | 0.027 | |
| 72. Brazil | 0.513 | Actual | 0.561 | 3.97 | 0.197 | -16.19 | 0.127 | -3.36 | 0.079 | 17.66 | 0.036 | 34.39 |
| | | Fair | 0.540 | | 0.235 | | 0.131 | | 0.067 | | 0.027 | |

Note: The Gini index and income shares are shown in decimals.

*(Continued).*



**Table A1. Actual income shares by quintile vs. fair income shares by quintile and the Gini index of 75 countries**. (Continued).

| Country | Gini index | Income share | Top 20% | %Δ | Fourth 20% | %Δ | Third 20% | %Δ | Second 20% | %Δ | Lowest 20% | %Δ |
|---|---|---|---|---|---|---|---|---|---|---|---|---|
| 73. Botswana | 0.533 | Actual | 0.585 | 4.00 | 0.195 | -14.25 | 0.111 | -10.81 | 0.070 | 14.18 | 0.039 | 60.19 |
| | | Fair | 0.562 | | 0.227 | | 0.124 | | 0.061 | | 0.024 | |
| 74. Zambia | 0.571 | Actual | 0.613 | 0.93 | 0.193 | -8.42 | 0.106 | -4.42 | 0.060 | 18.03 | 0.029 | 43.80 |
| | | Fair | 0.607 | | 0.211 | | 0.111 | | 0.051 | | 0.020 | |
| 75. Namibia | 0.591 | Actual | 0.637 | 0.87 | 0.179 | -10.92 | 0.098 | -5.46 | 0.058 | 26.97 | 0.028 | 53.95 |
| | | Fair | 0.632 | | 0.201 | | 0.104 | | 0.046 | | 0.018 | |

Note: The Gini index and income shares are shown in decimals.